\documentclass[preprint,showpacs,preprintnumbers,amsmath,amssymb,showkeys]{revtex4}
\def\texpsfig#1#2#3{\vbox{\kern #3\hbox{\includegraphics{#1}\kern #2}}\typeout{(#1)}}

\input epsf   

\usepackage{graphicx}
\usepackage{dcolumn}
\usepackage{bm}

\begin{document}

\title{A Variational Perturbation Approach \\ to One-Point Functions in QFT}

\author{Wen-Fa Lu}
   \email{wenfalu@sjtu.edu.cn}
\affiliation{ \ Institute for Theoretical Physics, Department of
Physics, Shanghai Jiao Tong University, Shanghai 200030, the
People's Republic of China \ (home institute)} \affiliation{ \ The
Abdus Salam International Centre for Theoretical Physics, Strada
Costiera, II 34014, Trieste, Italy}
\date{\today}

\begin{abstract}
In this paper, we develop a variational perturbation (VP) scheme
for calculating vacuum expectation values (VEVs) of local fields
in quantum field theories. For a comparatively general scalar
field model, the VEV of a comparatively general local field is
expanded and truncated at second order in the VP scheme. The
resultant truncated expressions (we call Gaussian smearing
formulae) consist mainly of Gaussian transforms of the local-field
function, the model-potential function and their derivatives, and
so can be used to skip calculations on path integrals in a
concrete theory. As an application, the VP expansion series of the
VEV of a local exponential field in the sine- and sinh-Gordon
field theories is truncated and derived up to second order
equivalently by directly performing the VP scheme, by finishing
ordinary integrations in the Gaussian smearing formulae, and by
borrowing Feynman diagrammatic technique, respectively.
Furthermore, the one-order VP results of the VEV in the
two-dimensional sine- and sinh-Gordon field theories are
numerically calculated and compared with the exact results
conjectured by Lukyanov, Zamolodchikov $et\; al.$, or with the
one-order perturbative results obtained by Poghossian. The
comparisons provide a strong support to the conjectured exact
formulae and illustrate non-perturbability of the VP scheme.
\end{abstract}


 \maketitle

\section{Introduction}
\label{1}

Green's functions and correlation functions in quantum field
theory(QFT), statistical mechanics and condensed matter physics
are closely related to experimental studies on macroscopic matter
systems and elementary particles \cite{1}. Because of the
existence of an operator-product-expansion algebra, various
multi-point Green's or correlation functions with the points
approaching each other can be reduced down to one-point functions
or vacuum expectation values (VEVs) of local fields \cite{2}. On
the other hand, as far as VEVs themselves go, they determine the
linear responses of statistical-mechanics systems to external
fields, and contain non-perturbative information about QFT, which
is not accessible through a direct investigation in the conformal
perturbation theory. Hence the problem of calculating the VEVs of
local fields is of fundamental significance.

There exist some methods for calculating the VEVs of local fields.
For an integrable QFT which can be considered as a conformal field
theory perturbed by some operator, the VEV of the perturbing
operator can be exactly obtained with the help of the
thermodynamic Bethe Ansatz approach \cite{3,4}. This kind of exact
calculation succeeded only in very few cases \cite{4}.
Nevertheless, S. Lukyanov and A. Zamolodchikov made a significant
progress in 1997 and conjectured the exact VEV of an exponential
field in the two-dimensional sine-Gordon (sG) and sinh-Gordon
(shG) field theories \cite{4}. This progress has led to the use of
reflection relation \cite{5} and accordingly exact VEVs of local
fields in many perturbed conformal field theories without and with
boundaries have been proposed \cite{6,7} through solving the
reflection relations. Furthermore, these conjectured exact VEVs
have again been employed to give VEVs of local fields in some
other QFTs by making use of quantum group restriction existed
between the relevant QFTs \cite{6,8}. Besides, recently, because
of the peculiarity in super-Liouville theories with boundary,
conformal bootstrap method \cite{9} and Modular transformation
method \cite{10} were used to derive one-point functions of bulk
and boundary operators. For the boundary scaling Lee-Yang model,
one-point functions of bulk and boundary fields were approximated
by using the truncated conformal space approach and the
form-factor expansion \cite{11}. Additionally, in order to check
those exact predictions, perturbation theory \cite{12} and some
numerical methods based on the truncated conformal space approach
\cite{13} were adopted to calculate the VEVs.

Obviously, since VEVs of local fields in QFT are non-perturbative
objects, general and systematical non-perturbation methods of
directly calculating VEVs are needed and such tools are also
necessary for independently checking those predictions on the
exact VEVs. We feel that a variational perturbation (VP) theory
\cite{14} can afford such a tool. The VP theory can be regarded as
some mixture of conventional perturbation theory and variational
method. As is well known, the perturbation theory is a
systematical approximation tool and results from it can be
improved order by order. Meanwhile, the variational method is
feasible and effective as well as valid for any coupling (the weak
or the strong). These two methods have been dominating
approximation calculations in theoretical researches for a long
time, being two standard approximation tools. However, the
perturbation theory is valid only for very weak coupling at most
and the variational method gives no indication of the error in its
resultant value. In order to collect their merits, avoid their
drawbacks, and, of course, to develop a systematical
non-perturbative tool, a primitive idea of the VP theory for
simply combining the conventional perturbation theory with the
variational method was proposed as a tool of solving stationary
Schr\"{o}dinger equation in 1955 \cite{15} (which can even date
back to even twenty more years earlier \cite{16}). This naive
combination has been applied to many fields in physics (see
references in Ref.~\cite{14,17,18}). It amounts to an expansion
around variational approximate result and the variational
parameter in the expansion is determined with the one-order result
by variational method. This primitive VP theory really produces
non-perturbative results which are valid for any coupling strength
and can be improved order by order. Nevertheless, the naive
combination of the perturbative and variational methods does not
provide a convergent tool because the variational parameter is
independent of the approximate order \cite{14}. In 1981 or so, the
principle of minimal sensitivity (PMS) \cite{19} was proposed by
P. M. Steveson (Possibly, earlier in the middle 1970s, a similar
principle was proposed in Mosc. Univ. Phys. Bull. {\bf 31},
10(1976) by V. I. Yukalov) and can be used to determine an
auxiliary parameter (the aforementioned variational parameter)
which is artificially introduced into the VP theory. The VP theory
with the PMS determines the parameter order by order (see next
section) and is believed to be a fast convergent theory \cite{14}.
Now, as a systematical and non-perturbative tool, it has developed
with many equivalent practical schemes to calculate energies, free
energies and effective potentials of systems, and has been applied
to QFT, condensed matter physics, statistical mechanics, chemical
physics, and so on \cite{14,17,18} (for a full list, to see
references in Refs.~\cite{14,17,18}). In this paper, we intend to
develop a VP scheme to calculate the VEVs of local fields in QFT.

In sect.II, we will develop the VP scheme in a general way. For
universality and definiteness, we shall consider a class of
systems, scalar field systems or Fermi field systems which can be
bosonized, with the Lagrangian density \cite{20}
\begin{equation}
{\cal L}_x={\frac {1}{2}}\partial_\mu \phi_x \partial^\mu \phi_x
-V(\phi_x) \;.
\end{equation}
In Eq.(1), the subscript $x=(\vec{x},t)$ represents the
coordinates in a $(D+1)$-dimensional Minkowski space,
$\partial_\mu$ and $\partial^\mu$ are the corresponding covariant
derivatives, and $\phi_x\equiv \phi(x)$ the scalar field at $x$.
The potential $V(\phi_x)$ in Eq.(1) is assumed to have a Fourier
representation in a sense of tempered distributions \cite{21},
\begin{equation}
V(\phi_x)=\int {\frac {d\omega}{\sqrt{2\pi}}} \tilde
{V}(\omega)e^{i\omega\phi_x} \;.
\end{equation}
Speaking roughly, this requires that the integral
$\int^\infty_{-\infty}V(\alpha) e^{-C\alpha^2}d\alpha$ with a
positive constant $C$ is finite. This shouldn't be regarded as a
limitation, and, as a matter of fact, quite a number of model
potentials possess the property, such as the potentials of
polynomial models, sG and shG models, Bullough-Dodd model,
Liouville model, two models discussed in Ref.~\cite{22}, and so
on. By the way, a similar general model was studied in
Ref.~\cite{2}. For comparisons with existed work, we will work in
a $\nu$-dimensional Euclidean space with $\nu\equiv (D+1)$.
Through the time continuation $t\to -i\tau$ with $\tau$ the
Euclidean time, a point $x$ in the Minkowski space is transformed
into a point $\vec{r}=(\vec{x},\tau)$ in the Euclidean space. For
universality and definiteness again, we will also assume that the
local field has a Fourier representation,
\begin{equation}
{\cal O}(\phi_{\vec{r}_0})=\int {\frac {d\Omega}{\sqrt{2\pi}}}
{\tilde {\cal O}}(\Omega)e^{i\Omega\phi_{\vec{r}_0}} \;,
\end{equation}
at least in a sense of tempered distributions. Here,
$\vec{r}_0=(\vec{x}_0,\tau_0)$ is a given point in the Euclidean
space. It is evident that such a local field is a comparatively
general one. In the present paper, based on the VP scheme in
Ref.~\cite{18} (the scheme stemmed from the Okopinska's optimized
expansion \cite{23}, and was proposed by Stancu and Stevenson
\cite{24}), we will develop a VP scheme to calculate the VEV of
the local field ${\cal O}(\phi_{\vec {r}_0})$, Eq.(3), for the
field theory Eq.(1) in the $\nu$-dimensional Euclidean space. In
subsection $A$ of next section, we will state the VP scheme. It
embraces mainly two key steps: one is the VP expansion on the VEVs
with an auxiliary parameter introduced, and another is the
determination of the auxiliary parameter in the expressions
truncated from the VP expansion series. It is the truncated
expressions that give rise to the VP approximate results of VEVs
up to the truncated order. Then, in subsection $B$ of next
section, the truncated expansions of the VEV of ${\cal
O}(\phi_{\vec {r}_0})$ in the field theory, Eq.(1), will
concretely be derived up to the second order. One will see that
the resultant truncated expressions up to the second order are
composed mainly of Gaussian transforms of the local field ${\cal
O}(\phi)$ and the potential $V(\phi)$ as well as their
derivatives, and we will call them Gaussian smearing formulae.
For, at least, any scalar field theory which is involved in the
model with Eq.(1), one can give VP approximated VEVs of ${\cal
O}(\phi_{\vec {r}_0})$ up to the second order just by finishing
ordinary integrations appeared in the Gaussian smearing formulae
instead of calculating path integrals in the definitions of the
VEVs. This point is the main reason why we are interested in the
general model, Eq.(1) and the general local fields, Eq.(3). We
should point out that since the QFT with Eq.(1) is not a concrete
theory, so we will not carry out the other key step to determine
the adjustable parameter in the Gaussian smearing formulae.
Section IV will provide such an example on how to determine the
auxiliary parameter by the PMS.

About the VP scheme, the renormalization problem need to be
explained here. Since a bare field theory is full of divergences
and makes no senses, we have to face those divergences appeared in
the Gaussian smearing formulae and perform a necessary
renormalization procedure to make the formulae finite before
performing the second key step with the PMS in the VP scheme.
Generally, the renormalization procedure is usually very
complicated, and is similar to those in perturbative theory
\cite{24}. In order to concentrate our attention at developing the
VP theory, we do not hope to be plagued with the complicated
renormalization problems of QFT, but we certainly hope to give a
finite example for the Gaussian smearing formulae. Fortunately,
for any two-dimensional scalar field theory with non-derivative
interactions, all ultraviolet divergences can be removed by
normal-ordering the Hamiltonian \cite{25,2}. This fact led to a
simple renormalization scheme for two-dimensional field theory,
the Coleman's normal-ordering renormalization prescription.
Furthermore, this convenient renormalization prescription has been
generalized to path integrals of Minkowski and Euclidean
formalisms in Ref.~\cite{25a} and Ref.~\cite{20} (2002),
respectively. Hence, this prescription was used in Ref.~\cite{18}
and will be adopted in the present paper. One will see that the
Gaussian smearing formulae are full of no explicit divergences for
the case of $\nu=2$ and so, generally, no further renormalization
procedures are needed for this case.

As an application of the scheme, Sections III and IV will
calculate the VEV of the local exponential field
$e^{ia\phi_{\vec{r}_0}}$ in the sG field theory, $G_a$, with $a$ a
parameter. The sG model, which appeared early in 1909 \cite{26},
is closely related to lots of problems in mathematics and physics,
and has been extensively studied. For the general local field
${\cal O}(\phi_{\vec{r}_0})$, Eq.(3), its VEV can closely be
related to that of an local exponential field, and so the problem
of calculating the VEVs of the local exponential field is
important. In 1997, through direct calculations in the sG field
theory at the coupling $\beta^2\to 0$ ( semi-classical limit) and
in the free-fermion theory (equivalent version of the sG field
theory at $\beta^2={\frac {1}{2}}$), and through the exact
specific free energy for the sG field theory, Lukyanov and
Zamolodchikov obtained the exact $G_a$ for the following three
special cases: $\beta\to 0$ (semi-classical limit), $\beta={\frac
{1}{2}}$ and $a=\beta$, respectively \cite{4}. And then, in the
same paper, starting from those exact expressions for the special
cases, they guessed an exact formula for $G_a$ in the
two-dimensional sG field theory at any $\beta^2<1$ and
$|Re(a)|<1/(2\beta)$. Obviously, it is worthwhile checking the
conjectured exact formula. In order to do so, in the same paper,
defining ``fully connected" one-point functions, $\sigma_{2n}$,
from the VEVs of even-power fields $\phi^{2n}$, they showed that
$\sigma_{2}$ and $\sigma_{4}$ from the above-mentioned exact
formula agree with those from perturbation theory for the sG field
theory up to $\beta^4$ and that $\sigma_{2}$ agrees with the
corresponding one-point function from perturbation theory up to
$g$ the coupling in the massive Thirring model, which is the
fermion version of the sG field theory \cite{25}(1975).
Furthermore, it was found that the conjectured exact formula can
be reobtained from the reflection relations \cite{5}, supporting
it indirectly. Slightly later, in 2000, checks from perturbation
theories in both an angular and a radial quantization approaches
for the massive Thirring model indicated that the perturbation
result up to $g$ exactly coincides with the corresponding result
obtained by expanding the exact formula according to the coupling
$g$ \cite{12}(2000). In the same year, a numerical study for the
model at a finite volume also provides evidence for the
conjectured exact formula \cite{13}(2000). In brief, up to now,
the conjectured exact formula for $G_a$ in the two-dimensional sG
field theory has been completely checked for the case of
$\beta^2\to {\frac {1}{2}}$ ($g\to 0$ is equivalent to $\beta^2\to
{\frac {1}{2}}$, See Eq.(67)), and received some indirect
evidences for its validity. So, besides providing an example of
applying the VP scheme, investigations on $G_a$ will also give the
conjectured formula a direct check for the cases of
$\beta^2\not={\frac {1}{2}}$ (Ref.~\cite{4} has provided a partial
check for $\beta\to 0$).

In section III, we will derive the expressions truncated from the
VP expansion series on $G_a$ at the second order. It will be done
in three ways. Subsection $A$ will directly perform the VP
expansion on $G_a$ and then truncate the expansion series at the
second order. Since the sG and shG field theories are involved in
the class of QFTs, Eq.(1), so, in subsection $B$, the truncated
expressions of $G_a$ up to second order will be recalculated by
finishing those ordinary integrations existed in the Gaussian
smearing formulae in Section II. The results are identical to
those obtained in subsection $A$. Thus, these calculations check
and confirm the correctness, usefulness and simplicity of the
Gaussian smearing formulae in Section II. Besides, the VP
expansion procedure is formally similar to conventional
perturbative expansion, and so the Feynman diagrammatic techniques
for perturbative theory, which have developed very well up to now,
can be borrowed into the VP expansion. Subsection $C$ will provide
such a calculation on $G_a$ up to the second order in the VP
expansion, and the resultant expressions are same as in Subsection
$A$ and $B$.

The resultant expressions in Sections II and III are not the final
VP approximate results because the other crucial step of the VP
scheme is not yet performed to determine the auxiliary parameter
by the PMS. As was aforementioned, Sect. IV will do it by
considering the truncated expressions up to the first order in
Sect. III and treat only the case of $\nu=2$. The sG field theory
can be transformed into the shG field theory, and so the one-order
VP results on the VEVs of the exponential fields in the
two-dimensional both sG and shG field theories will numerically
calculated and be completely compared with the conjectured exact
results. For the sG model, we will also compare the one-order VP
results here with those from a perturbation theory in
Ref.~\cite{12}(Poghossian). These comparisons, albeit just the
one-order results are considered, not only give a strong support
for the conjectured exact results but also indicate usefulness and
non-perturbability of the VP scheme here for calculating one-point
functions. By the way, the VP approximate result up to the second
order on the VEV of the exponential fields in the two-dimensional
sG field theories has briefly been reported in Ref.~\cite{27}, and
has a less error to the conjectured exact results than the
one-order result. This gives a sign for the convergency of the VP
scheme here.

Briefly, next section, we will develop the VP scheme by
calculating the VEV's of the local field, Eq.(3), in the QFT,
Eq.(1), and Sect. III and IV will provide an application of the
scheme to the sG and shG field theories. A brief conclusion will
be made in Sect. V.

\section{The VP Approach to VEVs of Local Fields in QFT}
\label{2}

As was stated in the introduction, we will first state the VP
scheme in subsection $A$, and then derive the first three terms in
the VP expansion series in subsection $B$.

\subsection{the VP Scheme}
\label{vpt}

The generating functional is usually the start of discussing a QFT
and will be the basis for the purpose in the present paper. In
this subsection, we will first introduce it, then state our VP
scheme.

For any field theory, one can use either the Minkowskian formalism
or Euclidean formalism to study it \cite{28}. Here, we choose to
have our discussion in Euclidean formalism. The Euclidean
Lagrangian corresponding to the Minkowskian Lagrangian, Eq.(1), is
\begin{equation}
{\cal L}_{\vec{r}}={\frac {1}{2}}\nabla_{\vec{r}}\phi_{\vec{r}}
\nabla_{\vec{r}} \phi_{\vec{r}} + V(\phi_{\vec{r}}) \;,
\end{equation}
which has the same form with the Hamiltonian density of Eq.(1) in
$\nu+1$ spacetime. In Euclidean formalism, the corresponding
generating functional takes the following form \cite{2}:
\begin{equation}
Z[J]=\int {\cal D}\phi \exp\{-\int d^\nu \vec{r} \ [
       {\frac {1}{2}}\nabla_{\vec{r}}\phi_{\vec{r}}
\nabla_{\vec{r}} \phi_{\vec{r}}+ V(\phi_{\vec{r}})
-J_{\vec{r}}\phi_{\vec{r}}\;]\} \; ,
\end{equation}
where, $\nabla_{\vec{r}}$ is the gradient in the $\nu$-dimensional
Euclidean space, $J_{\vec{r}}$ an external source at $\vec{r}$,
and ${\cal D}\phi$ the functional measure. Generally, for an
interacting system, the right side hand of Eq.(5) can not exactly
be calculated. But for a free field system with
$V(\phi_{\vec{r}})={\frac {1}{2}} \mu^2 \phi_{\vec{r}}^2 $, its
generating functional is exactly calculable and plays a crucial
role in the VP theory and relevant calculations. For the
convenience of later uses, we write down its result as follows
\cite{2}
\begin{eqnarray}
Z_\mu[J]&=&\int {\cal D}\phi \exp\{-\int d^\nu \vec{r} \ [\ {\frac
{1}{2}}\nabla_{\vec{r}} \phi_{\vec{r}}\nabla_{\vec{r}}
\phi_{\vec{r}} + {\frac {1}{2}} \mu^2 \phi_{\vec{r}}^2
-J_{\vec{r}}\phi_{\vec{r}}\ ]\} \nonumber \\
&=&\exp\{-{\frac {1}{2}}\int d^\nu \vec{r}\;
I_{(0)}(\mu^2)\}\exp\{{\frac {1}{2}}Jf^{-1}J\}\;.
\end{eqnarray}
Here, \begin{eqnarray*} I_{(n)}(Q^2)\equiv\Biggl \{
    \begin{array}{ll}
\int {\frac {d^\nu \vec{p}}{(2\pi)^2}}
{\frac {1}{(p^2+Q^2)^n}} \;, & \ \ \ \ \ {\rm for} \ \ n\not=0  \\
 \int {\frac {d^\nu \vec{p}}{(2\pi)^2}} \ln(p^2+Q^2)  \;, & \ \ \ \ \
                          {\rm for} \ \  n=0
    \end{array}
\end{eqnarray*}
with $\vec{p}$ momentum and $p=|\vec{p}|$  and $Jf^{-1}J\equiv\int
d^\nu \vec{r}' d^\nu \vec{r}'' J_{\vec{r}'}f^{-1}_{\vec{r}'
\vec{r}''}J_{\vec{r}''}$ with \cite{24}(1990)
\begin{equation}
f^{-1}_{\vec{r}' \vec{r}''}\equiv \int {\frac {d^\nu
\vec{p}}{(2\pi)^\nu}} {\frac {1}{p^2+\mu^2}} \; e^{i\vec{p}\cdot
(\vec{r}''-\vec{r}')}={\frac
{\mu^{\nu/2-1}}{(2\pi)^{\nu/2}|\vec{r}''-\vec{r}'|^{\nu/2-1}}}K_{\nu/2-1}(\mu|\vec{r}''-\vec{r}'|)
\;.
\end{equation}
In Eq.(7), $K_n(z)$ is the $n$th-order modified Bessel function of
the second kind.

Now we address ourselves to the VEVs. For a local field ${\cal
O}(\phi_{\vec {r}_0})$, its VEV in the theory with Eq.(4) is
defined as follows \cite{4}
\begin{equation}
{\bar {\cal O}}\equiv <{\cal O}(\phi_{\vec {r}_0})>\equiv {\frac
{\int {\cal D}\phi {\cal O}(\phi_{\vec {r}_0}) \exp\{-\int d^\nu
\vec{r}{\cal L}_{\vec {r}}\}} {\int {\cal D}\phi \exp\{-\int d^2
\vec{r}{\cal L}_{\vec {r}}\}}}  \;.
\end{equation}
In general, ${\bar {\cal O}}$ in last equation cannot be exactly
calculated and one has to manage to design some approximate tool
to attack it.

According to Eqs.(3),(4) and (5), ${\bar {\cal O}}$ in Eq.(8) can
easily be rewritten as
\begin{equation}
{\bar {\cal O}}=\int {\frac {d\Omega}{\sqrt{2\pi}}} {\tilde {\cal
O}}(\Omega){\frac {Z[J]_{J_{\vec{r}}= i\Omega
\delta(\vec{r}-\vec{r}_0)}} {Z[J=0]}} \;.
\end{equation}
In the numerator of the integrand in Eq.(9), the subscript means
that $i\Omega \delta(\vec{r}-\vec{r}_0)$ takes the place of
$J_{\vec{r}}$ in $Z[J]$. Hence, ${\bar {\cal O}}$ can be obtained
via calculating the generating functional. Luckily, the
calculation of effective potential has stimulated the
establishment of several VP expansion schemes on the generating
functional (see Refs.~\cite{14,17} and references therein), and
they can possibly be developed to calculate ${\bar {\cal O}}$.
Next, we will generalize the VP scheme in the Ref.~\cite{18}(a
very slightly different version of that in Ref.~\cite{24}) to
calculate ${\bar {\cal O}}$.

As was stated in the introduction, the Coleman's normal-ordering
renormalization prescription will be adopted. That is, the
Euclidean Lagrangian in the exponential in Eq.(5) will be replaced
by the following normal-ordered form with respect to an arbitrary
normal-ordering mass $\cal M$ \cite{20} (2002),
$$-{\frac {1}{2}}I_{(0)}({\cal M}^2)
       +{\frac {1}{2}} {\cal M}^2 I_{(1)}({\cal M}^2)+{\frac {1}{2}}\nabla_{\vec{r}}\phi_{\vec{r}}
\nabla_{\vec{r}} \phi_{\vec{r}}
        -J_{\vec{r}}\phi_{\vec{r}}
        +\int {\frac {d \omega}{\sqrt{2\pi}}}\tilde{V}(\omega)
           e^{i \omega\phi_{\vec{r}}} e^{{\frac {1}{2}}\omega^2 I_{(1)}({\cal M}^2)}\;.$$
In the above expression, the last term is the normal-ordered form
of $V(\phi_{\vec{r}})$, and can be got by using Eq.(2) and the
Baker-Hausdorff formula $e^{A+B}=e^A e^B e^{-[A,B]}$ with the
commutator $[A,B]$ some c-number. So, $Z[J]$ is turned into the
following form :
\begin{eqnarray}
Z[J]&=&\exp\{\int d^\nu r [{\frac {1}{2}}I_{(0)}({\cal M}^2)
        -{\frac {1}{2}} {\cal M}^2 I_{(1)}({\cal M}^2)]\}\int {\cal D}\phi
 \nonumber  \\  & \ \ \ & \times
 \exp\{-\int d^\nu \vec{r} \ [
       {\frac {1}{2}}\nabla_{\vec{r}}\phi_{\vec{r}}
\nabla_{\vec{r}} \phi_{\vec{r}}
        -J_{\vec{r}}\phi_{\vec{r}}
        +\int {\frac {d \omega}{\sqrt{2\pi}}}\tilde{V}(\omega)
           e^{i \omega\phi_{\vec{r}}} e^{{\frac {1}{2}}\omega^2 I_{(1)}({\cal M}^2)}
        ]\}  \;.
\end{eqnarray}
Eq.(10) is nothing but Eq.(11) of Ref.~\cite{20} (2002) in
notations here. Thus, for the case of two dimensions, the fields
and the model parameters, such as mass and couplings, are now
finite \cite{25,2}(for simplicity, we use the same symbols as
those in the bare Lagrangian, Eq.(4)).

Now we further modify $Z[J]$ in Eq.(10) by following
Ref.~\cite{18} or \cite{20}(2002)(only without shifting the field,
for simplicity). First, a parameter $\mu$ is introduced by adding
a vanishing term$\int d^\nu \vec{r}{\frac
{1}{2}}\phi_{\vec{r}}(\mu^2-\mu^2)\phi_{\vec{r}}$ into the
exponent of the functional integral in Eq.(10). This way of
introducing the artificial $\mu$ was used several decades ago
\cite{28b}. Then, rearrange the exponent into a free-field part
(with a mass being the introduced parameter $\mu$) plus a new
interacting part. Finally insert a formal expansion factor
$\epsilon$ in front of the interacting part. Consequently, $Z[J]$
is turned into the following form
\begin{eqnarray}
Z[J;\epsilon]&=&\exp\{\int d^\nu \vec{r}\; [{\frac
      {1}{2}}I_{(0)}[M^2]-{\frac {1}{2}} M^2 I_{(1)}[M^2]]\}
      \nonumber  \\   &\ \ \ & \times \int {\cal D}\phi
      \exp\{-\int d^\nu \vec{r}
      [{\frac {1}{2}}\phi_{\vec{r}}(-\nabla^2_{\vec{r}}+\mu^2)
      \phi_{\vec{r}}+\epsilon  {\cal H}_I
      (\phi_{\vec{r}},\mu)-J_r\phi_{\vec{r}}]\}
        \nonumber\\
        &=&\exp\{\int d^\nu \vec{r}\; [{\frac
      {1}{2}}I_{(0)}[M^2]-{\frac {1}{2}} M^2 I_{(1)}[M^2]]\}
      \nonumber  \\   &\ \ \ & \times \int {\cal D}\phi
      \exp\{-\epsilon \int d^\nu \vec{r}\; {\cal H}_I
      ({\frac {\delta}{\delta J_{\vec{r}}}},\mu)\}
      \exp\{-\int d^\nu \vec{r}
      [{\frac {1}{2}}\phi_{\vec{r}}(-\nabla^2_{\vec{r}}+\mu^2)
      \phi_{\vec{r}}-J_r\phi_{\vec{r}}]\}
        \nonumber\\
        &=&\exp\{-\int d^\nu \vec{r}\; [{\frac
      {1}{2}} I_{(0)}(\mu^2)-{\frac {1}{2}}I_{(0)}({\cal M}^2)+{\frac
      {1}{2}}{\cal M}^2 I_{(1)}({\cal M}^2)]\} \nonumber
      \\ && \times \exp\{-\epsilon\int d^\nu \vec{r}\; {\cal H}_I
      ({\frac {\delta}{\delta J_{\vec{r}}}},\mu)\}
      \exp\{{\frac {1}{2}}Jf^{-1}J\} \;
\end{eqnarray}
with the new interacting part
\begin{equation}
{\cal H}_I(\phi_{\vec{r}},\mu)=-{\frac
{1}{2}}\mu^2\phi_{\vec{r}}^2+\int {\frac {d
\omega}{\sqrt{2\pi}}}\tilde{V}(\omega)
           e^{i \omega\phi_{\vec{r}}} e^{{\frac {1}{2}}\omega^2 I_{(1)}({\cal M}^2)} \;.
\end{equation}
In writing down Eq.(11), we have employed the result Eq.(6). Note
that extrapolating $Z[J;\epsilon]$ to $\epsilon=1$, one recovers
$Z[J]$ in Eq.(10).

Expanding $\exp\{-\epsilon\int d^\nu \vec{r}\; {\cal H}_I({\frac
{\delta}{\delta J_{\vec{r}}}},\mu)\}$ in Eq.(11) into a series in
$\epsilon$, one has
\begin{eqnarray}
Z[J;\epsilon]&=&\exp\{-\int d^\nu \vec{r}\; [{\frac {1}{2}}
I_{(0)}(\mu^2)-{\frac {1}{2}}I_{(0)}({\cal M}^2)+{\frac {1}{2}}{\cal
M}^2 I_{(1)}({\cal M}^2)]\} \nonumber
\\ && \times  \sum_{n=0}^\infty \epsilon^n {\frac {(-1)^{n}}{n!}} \int \prod_{k=1}^n
d^\nu \vec{r}_k {\cal H}_I({\frac {\delta}{\delta
J_{\vec{r}_k}}},\mu)\exp\{{\frac {1}{2}}Jf^{-1}J\} \; .
\end{eqnarray}
Thus, we obtain an expansion of the generating functional without
requiring the model coupling and/or the formal expansion factor
$\epsilon$ very weak. It seems that this expansion form has no
senses and is useless. Nevertheless, when the PMS is introduced,
it can really furnish quite a potential non-perturbative
approximate tool for us to attack physical problems
\cite{14,17,18}. Next, one will see that it provides a basis for
non-perturbatively producing approximate results of ${\bar {\cal
O}}$.

Since we are using normal-ordered form, the local field ${\cal
O}(\phi_{\vec {r}_0})$ should be replaced by its normal-ordered
form from now on, and, from Eq.(3) as well as the Baker-Hausdorff
formula, it can be written as
$${\cal N}_{\cal M}[{\cal O}(\phi_{\vec{r}_0})]=\int {\frac {d\Omega}{\sqrt{2\pi}}} {\tilde
{\cal O}}(\Omega)\exp\{{\frac {1}{2}}\Omega^2 I_{(1)}({\cal
M}^2)\}e^{i\Omega(\phi_{\vec{r}_0})}$$ at least in a sense of
tempered distributions. Then ${\bar {\cal O}}$ in Eq.(9) can take
the following form
\begin{eqnarray}
{\bar {\cal O}}&=&\int {\frac {d\Omega}{\sqrt{2\pi}}} {\tilde {\cal
O}}(\Omega)\exp\{{\frac {1}{2}}\Omega^2 I_{(1)}({\cal M}^2)\}
 \nonumber \\
&& \times {\frac {\bigl[\sum_{n=0}^\infty \epsilon^n {\frac
{(-1)^{n}}{n!}} \int \prod_{k=1}^n d^\nu \vec{r}_k {\cal H}_I({\frac
{\delta}{\delta J_{\vec{r}_k}}},\mu)\exp\{{\frac
{1}{2}}Jf^{-1}J\}\bigl]_{J_{\vec{r}}= i\Omega
\delta(\vec{r}-\vec{r}_0)}}{\bigl[\sum_{n=0}^\infty \epsilon^n
{\frac {(-1)^{n}}{n!}} \int \prod_{k=1}^n d^\nu \vec{r}_k {\cal
H}_I({\frac {\delta}{\delta J_{\vec{r}_k}}},\mu)
     \exp\{{\frac {1}{2}}Jf^{-1}J\}\bigl]_{J=0}}}\Biggl|_{\epsilon=1} \;.
\end{eqnarray}
The right hand side of Eq.(14) is only a quotient between two
series in terms of $\epsilon$ and can be rearranged into a series
consisting of powers of $\epsilon$
\begin{equation}
{\bar {\cal O}}=[{\bar {\cal O}}^{(0)}+\epsilon^1 {\bar {\cal
O}}^{(1)}+\epsilon^2 {\bar {\cal O}}^{(2)}+\cdots +\epsilon^n
{\bar {\cal O}}^{(n)} +\cdots ]_{\epsilon=1}=
\Bigl[\sum_{n=0}^\infty \epsilon^n {\bar {\cal
O}}^{(n)}\Bigl]_{\epsilon=1} \; .
\end{equation}
This can be realized according to the formula 0.313 on page 14 in
Ref.~\cite{29},
\begin{equation*}
{\frac {\sum_{n=0}^\infty b_n z^n}{\sum_{n=0}^\infty a_n z^n}} =
{\frac {1}{a_0}}\sum_{n=0}^\infty c_n z^n, \hspace*{0.6cm}
c_n+{\frac {1}{a_0}}\sum_{k=1}^n c_{n-k}a_k-b_n=0 \;.
\end{equation*}

This series, Eq.(15), is the VP expansion series with the
auxiliary parameter $\mu$. For $\nu=2$, every terms in the series
Eq.(15) must be full of no explicit divergences (there are
possibly some implied divergences for some theories, and such an
example will be met in next section), and truncating it at some
order of $\epsilon$ will lead to an approximate ${\bar {\cal O}}$.
In the case of $\nu>2$ where divergences are met in Eq.(15), the
model parameters (the mass, coupling constants) should be written
as series in $\epsilon$ as did in Ref.~\cite{24}, respectively,
and then one can successively do as follow: substitute the series
of the model parameters into the truncated expression, rearrange
the truncated expression in terms of powers in $\epsilon$, keep
terms in it only up to the truncated order. After these, one can
take $\epsilon=1$ in the truncated expression and manage to find a
renormalization scheme for rendering the truncated expression
explicitly finite (finding the renormalization scheme is similar
to what is done in a perturbative theory). The finite truncated
expression at any order will be dependent upon $\mu$. How can we
determine the arbitrary parameter $\mu$ in a truncated result? As
was stated in the introduction, we can determine $\mu$ with the
PMS \cite{19}. Evidently, the exact result, Eq.(15) with
$\epsilon=1$, is independent of the auxiliary parameter $\mu$ and
is a constant in the space of $\mu$. Hence, it would be reasonable
that the truncated result should vary most slowly with the
parameter $\mu$ so that it can likely provide a most reliable
approximate result to the exact one. This is the main spirit of
the so called PMS \cite{19}. A simple realization of the PMS is to
require the first derivative of the truncated expression with
respect to $\mu$ to be zero. Sometimes, it cannot give rise to a
meaningful solution, and in this case one can render the second
derivative zero to determine $\mu$. Thus the above procedure,
truncating the series in Eq.(15) at some order of $\epsilon$ and
using the PMS to determine $\mu$, provides an approximate method
of calculating ${\bar {\cal O}}$ which can systematically control
its approximate accuracy.

In a general, we can say nothing on the convergent property of
Eq.(15), let alone on the convergency of the sequence consisting
of the truncated results at various orders, because it is
difficult to prove their convergency in a general way.
Nevertheless, from the aforementioned spirit of the PMS, the
convergency of the sequence of the truncated results determined
with the PMS to the exact result should be conceivable and
understandable. Existed investigations and applications have
indicated and illustrated the point. In the early 1980s, Stevenson
proposed the PMS, studied and indicated the efficiency of the PMS
for keeping the VP theory convergent. In 1990s, it was rigorously
proved that it is owing to the PMS that the VP theory leads to
quickly convergent results on ground state energy for a
quantum-mechanical anharmonic oscillator \cite{30}\cite{14}(2004),
and recently, the rigorous proofs of the convergency on the VP
theory have been given to the critical O(N) scalar field theory
\cite{31}. From these existed investigations, the VP scheme of
calculating VEVs in the present paper is presumably convergent,
because, frankly, it is just a generalization of the existed VP
schemes of calculating energy, free energy, effective potentials,
and so on. As was aforementioned, a brief report in Ref.~\cite{27}
suggest a sign on the convergency of the VPT scheme here. In the
present paper, we will not discuss the convergent problem.

The VP scheme of calculating VEVs stated in the above is a
non-perturbative method. The expansion in Eq.(15) in terms of
$\epsilon$ is intrinsically different in nature from a
conventional perturbation expansion in terms of the interaction
coupling, albeit they are formally similar to each other. The key
point of the reliable approximate results from conventional
perturbation expansion consists in the requirement of weak
couplings, whereas the crucial point of the reliable and fast
convergent approximate results from the VP expansion consists in
the requirement of most insensitivity of the truncated results to
the auxiliary parameter. It is due to the adjustability of the
auxiliary parameter that the expansion in Eq.(15) has no
limitations both to the formal expansion factor $\epsilon$ (in
fact, $\epsilon=1$ is taken eventually) and to the coupling. In
the VP expansion, there exists the adjustable term, $-{\frac
{1}{2}}\mu^2\phi_{\vec{r}}^2$, besides the original interaction
terms (see Eq.(12)), and the propagator Eq.(7), which is embraced
in the unexpanded part of $Z[J;\epsilon]$, also contains the
adjustable parameter, $\mu$. So it is conceivable that for a
different value of the coupling (weak or strong), one can adjust
the value of $\mu$ through the PMS to get to a reliable
approximate result of ${\bar {\cal O}}$. The existed
investigations have indicated the non-perturbability of the VP
theory. In Section IV, we will provide a comparison of the VP
theory here with perturbation theory to illustrate the
non-perturbative property of the VP scheme here.

\subsection{Truncate the VP Expansion Series at the Second Order}
\label{gsf}

This subsection derives the first three terms in the series
Eq.(15). At the zeroth order, Eqs.(14) and (15) lead to
\begin{eqnarray}
{\bar {\cal O}}^{(0)}&=& \int {\frac {d\Omega}{\sqrt{2\pi}}} {\tilde
{\cal O}}(\Omega)\exp\{{\frac {1}{2}}\Omega^2 I_{(1)}({\cal
M}^2)\}\biggl[\exp\{{\frac
{1}{2}}Jf^{-1}J\}\biggl]_{J_{\vec{r}}=i\Omega
\delta(\vec{r}-\vec{r}_0)} \nonumber \\
&=& \int {\frac {d\Omega}{\sqrt{2\pi}}} {\tilde {\cal
O}}(\Omega)e^{-{\frac {1}{2}}\Omega^2 {\cal U}}=\int {\frac
{d\Omega}{\sqrt{2\pi}}} {\tilde {\cal
O}}(\Omega)\int_{-\infty}^{\infty} {\frac
{d\alpha}{\sqrt{2\pi}}}e^{-{\frac {\alpha^2}{2}}+i\Omega\alpha\sqrt{\cal U}} \nonumber \\
 &=& \int_{-\infty}^{\infty} {\frac
{d\alpha}{\sqrt{2\pi}}}e^{-{\frac {\alpha^2}{2}}}{\cal
O}(\alpha\sqrt{\cal U})
\end{eqnarray}
with ${\cal U}\equiv I_{(1)}(\mu^2)-I_{(1)}({\cal M}^2)$. Here, we
have used Eq.(3) and the Gaussian integral formula
$\int^{\infty}_{-\infty}{\frac {d\alpha}{\sqrt{2\pi}}}e^{-{\frac
{\alpha^2}{2}}+\alpha\sqrt{2 a}}=e^{a}$, which will repeatedly be
used for obtaining next Eqs.(20) and (25).

At the first order, the coefficient ${\bar {\cal O}}^{(1)}$ is as
follows
\begin{eqnarray}
{\bar {\cal O}}^{(1)}&=&-\int {\frac {d\Omega}{\sqrt{2\pi}}}
{\tilde {\cal O}}(\Omega)\exp\{{\frac {1}{2}}\Omega^2
I_{(1)}({\cal M}^2)\}\Biggl[\int d^\nu \vec{r}_1 {\cal H}_I({\frac
{\delta}{\delta J_{\vec{r}_1}}},\mu) \exp\{{\frac
{1}{2}}Jf^{-1}J\}\Biggl]_{J_{\vec{r}}=i\Omega
\delta(\vec{r}-\vec{r}_0)} \nonumber
\\ &&+{\bar {\cal O}}^{(0)}\Biggl[\int d^\nu \vec{r}_1 {\cal H}_I({\frac {\delta}{\delta
J_{\vec{r}_1}}},\mu)\exp\{{\frac
{1}{2}}Jf^{-1}J\}\Biggl]_{J_{\vec{r}}=0} \;.
\end{eqnarray}
To calculate ${\bar {\cal O}}^{(1)}$, one can first have
\begin{equation}
{\frac {\delta^2}{\delta J_{\vec{r}_1}^2}}\exp\{{\frac
{1}{2}}Jf^{-1}J\}=[f^{-1}_{\vec{r}_1
\vec{r}_1}+(Jf^{-1})_1^2]\exp\{{\frac {1}{2}}Jf^{-1}J\}
\end{equation}
and
\begin{equation}
\exp\{i\omega{\frac {\delta}{\delta J_{\vec{r}_1}}}\}\exp\{{\frac
{1}{2}}Jf^{-1}J\}=\exp\{{\frac
{1}{2}}Jf^{-1}J\}\Bigl|_{J_{\vec{r}}\to J_{\vec{r}}+ i\omega
\delta(\vec{r}-\vec{r}_1)}
\end{equation}
with $(Jf^{-1})_n^m \equiv (\int d^\nu \vec{r} J_{\vec{r}}
f^{-1}_{\vec{r}\vec{r_n}})^m$. Eq.(19) is easily obtained by
returning its left hand side to its original integral expression.
(So are Eqs.(23) and (24)). Then substituting them into Eq.(17),
one can obtain
\begin{eqnarray}
{\bar {\cal O}}^{(1)}&=&{\frac {1}{2}}\mu^2\int {\frac
{d\Omega}{\sqrt{2\pi}}} {\tilde {\cal O}}(\Omega)e^{-{\frac
{1}{2}}\Omega^2 {\cal U}}\int d^\nu \vec{r}_1 [f^{-1}_{\vec{r}_1
\vec{r}_1}+(i\Omega)^2(f^{-1}_{\vec{r}_0 \vec{r}_1})^2]
  \nonumber \\ &&-\int
{\frac {d\Omega}{\sqrt{2\pi}}}{\frac {d\omega}{\sqrt{2\pi}}}\int
d^\nu \vec{r}_1 {\tilde {\cal O}}(\Omega){\tilde V}(\omega)
e^{-{\frac {1}{2}}(\Omega^2+\omega^2) {\cal U}} e^{-\omega\Omega
f^{-1}_{\vec{r}_0 \vec{r}_1}}
  \nonumber \\ &&
-{\frac {1}{2}}\mu^2{\bar {\cal O}}^{(0)}\int d^\nu \vec{r}_1
f^{-1}_{\vec{r}_1 \vec{r}_1} +{\bar {\cal O}}^{(0)} \int {\frac
{d\omega}{\sqrt{2\pi}}}\int d^\nu \vec{r}_1 {\tilde V}(\omega)
e^{-{\frac {1}{2}}\omega^2 {\cal U}}
  \nonumber \\
  &=&{\frac
{1}{2}}\mu^2\int d^\nu \vec{r}_1 (f^{-1}_{\vec{r}_0 \vec{r}_1})^2
\int^\infty_{-\infty}{\frac {d\alpha}{\sqrt{2\pi}}}e^{-{\frac
{\alpha^2}{2}}}{\cal O}^{(2)}(\alpha\sqrt{\cal U}) \nonumber
 \\ && - \int d^\nu \vec{r}_1 \int^\infty_{-\infty}{\frac
{d\gamma}{\sqrt{2\pi}}}e^{-{\frac {\gamma^2}{2}}} V(\gamma
\sqrt{\cal U}) \nonumber  \\
&&  \hspace*{1cm}\times \int^\infty_{-\infty}{\frac
{d\alpha}{\sqrt{2\pi}}}e^{-{\frac {\alpha^2}{2}}}\Bigl[{\cal
O}(\alpha\sqrt{{\cal W}_{01}}+\gamma {\frac {f^{-1}_{\vec{r}_0
\vec{r}_1}}{\sqrt{\cal U}}})-{\cal O}(\alpha\sqrt{\cal U})\Bigl]
\;,
\end{eqnarray}
where ${\cal O}^{(n)}(z)={\frac {d^n {\cal O}(z)}{(d z)^n}}=\int
{\frac {d \Omega} {\sqrt{2\pi}}}\tilde{\cal O}(\Omega) (i
\Omega)^n e^{i \Omega z}$ and ${\cal W}_{jk}\equiv {\frac {{\cal
U}^2-(f^{-1}_{\vec{r}_j \vec{r}_k})^2}{{\cal U}}}$.

Finally, Eqs.(14) and (15) give the second-order coefficient in
Eq.(15) as
\begin{eqnarray}
{\bar {\cal O}}^{(2)}&=& \int {\frac {d\Omega}{\sqrt{2\pi}}} {\tilde
{\cal O}}(\Omega)\exp\{{\frac {1}{2}}\Omega^2 I_{(1)}({\cal
M}^2)\}{\frac {1}{2!}}\Bigl[\int d^\nu \vec{r}_1 d^\nu \vec{r}_2
{\cal H}_I({\frac {\delta}{\delta J_{\vec{r}_1}}},\mu)\nonumber \\
&& \hspace*{3cm}\times  {\cal H}_I({\frac {\delta}{\delta
J_{\vec{r}_2}}},\mu)\exp\{{\frac {1}{2}}Jf^{-1}J\}
\Bigl]_{J_{\vec{r}}=i\Omega
\delta(\vec{r}-\vec{r}_0)}\nonumber \\
&&-{\frac {1}{2!}}{\bar {\cal O}}^{(0)}\Bigl[\int d^\nu \vec{r}_1
d^\nu \vec{r}_2 {\cal H}_I({\frac {\delta}{\delta
J_{\vec{r}_1}}},\mu){\cal H}_I({\frac {\delta}{\delta
J_{\vec{r}_2}}},\mu)\exp\{{\frac
{1}{2}}Jf^{-1}J\} \Bigl]_{J_{\vec{r}}=0} \nonumber \\
&&+{\bar {\cal O}}^{(1)}\Bigl[\int d^\nu \vec{r}_1 {\cal H}_I({\frac
{\delta}{\delta J_{\vec{r}_1}}},\mu)\exp\{{\frac
{1}{2}}Jf^{-1}J\}\Bigl]_{J_{\vec{r}}=0}\;.
\end{eqnarray}
Similarly to what was done at the first order, one can first have
\begin{eqnarray}
{\frac {\delta^2}{\delta J_{\vec{r}_1}^2}}{\frac {\delta^2}{\delta
J_{\vec{r}_2}^2}}\exp\{{\frac {1}{2}}Jf^{-1}J\} &=& \{2
(f^{-1}_{\vec{r}_1 \vec{r}_2})^2+f^{-1}_{\vec{r}_1
\vec{r}_1}f^{-1}_{\vec{r}_2 \vec{r}_2}+ 4 f^{-1}_{\vec{r}_1
\vec{r}_2}(Jf^{-1})_1(Jf^{-1})_2+ f^{-1}_{\vec{r}_1
\vec{r}_1}(Jf^{-1})_2^2 \nonumber  \\ &&
\hspace*{0.5cm}+f^{-1}_{\vec{r}_2 \vec{r}_2}(Jf^{-1})_1^2 +
(Jf^{-1})_1^2(Jf^{-1})_2^2\}\exp\{{\frac {1}{2}}Jf^{-1}J\} \; ,
\end{eqnarray}
\begin{equation}
 {\frac {\delta^2}{\delta
J_{\vec{r}_1}^2}} \exp\{i\omega{\frac {\delta}{\delta
J_{\vec{r}_2}}}\}\exp\{{\frac
{1}{2}}Jf^{-1}J\}=\Bigl\{[f^{-1}_{\vec{r}_1
\vec{r}_1}+(Jf^{-1})_1^2]\exp\{{\frac
{1}{2}}Jf^{-1}J\}\Bigl\}\Bigl|_{J_{\vec{r}}\to J_{\vec{r}}+
i\omega \delta(\vec{r}-\vec{r}_2)}
\end{equation}
and
\begin{equation}
\exp\{i\omega_1{\frac {\delta}{\delta
J_{\vec{r}_1}}}\}\exp\{i\omega_2{\frac {\delta}{\delta
J_{\vec{r}_2}}}\}\exp\{{\frac {1}{2}}Jf^{-1}J\}=\exp\{{\frac
{1}{2}}Jf^{-1}J\}\Bigl|_{J_{\vec{r}}\to J_{\vec{r}}+ i\omega_1
\delta(\vec{r}-\vec{r}_1)+ i\omega_2 \delta(\vec{r}-\vec{r}_2)} ,
\end{equation}
and then substituting Eqs.(22), (23) and (24) into Eq.(21) leads
to
\begin{eqnarray}
{\bar {\cal O}}^{(2)}&=&{\frac {\mu^4}{8}}\int d^\nu \vec{r}_1
d^\nu \vec{r}_2 \int {\frac {d\Omega}{\sqrt{2\pi}}} {\tilde {\cal
O}}(\Omega)[2(f^{-1}_{\vec{r}_1 \vec{r}_2})^2+(f^{-1}_{\vec{r}_1
\vec{r}_1})^2+4(i\Omega)^2f^{-1}_{\vec{r}_0
\vec{r}_1}f^{-1}_{\vec{r}_0 \vec{r}_2}f^{-1}_{\vec{r}_1
\vec{r}_2}
 \nonumber \\ && \hspace*{5cm}
+2(i\Omega)^2 f^{-1}_{\vec{r}_1 \vec{r}_1} (f^{-1}_{\vec{r}_0
\vec{r}_1})^2+(i\Omega)^4(f^{-1}_{\vec{r}_0
\vec{r}_1})^2(f^{-1}_{\vec{r}_0 \vec{r}_2})^2]e^{-{\frac
{1}{2}}\Omega^2 {\cal U}}
 \nonumber \\ &&
-{\frac {\mu^2}{2}}\int d^\nu \vec{r}_1 d^\nu \vec{r}_2 \int
{\frac {d\Omega}{\sqrt{2\pi}}}{\frac {d\omega}{\sqrt{2\pi}}}
{\tilde {\cal O}}(\Omega){\tilde V}(\omega)[f^{-1}_{\vec{r}_1
\vec{r}_1}+(i\Omega)^2 (f^{-1}_{\vec{r}_0
\vec{r}_1})^2+(i\omega)^2(f^{-1}_{\vec{r}_1 \vec{r}_2})^2
 \nonumber \\ && \hspace*{5cm}
+2i\omega i\Omega f^{-1}_{\vec{r}_0 \vec{r}_1} f^{-1}_{\vec{r}_1
\vec{r}_2}]e^{-{\frac {1}{2}}(\Omega^2+\omega^2) {\cal
U}-\omega\Omega f^{-1}_{\vec{r}_0 \vec{r}_2}}
 \nonumber \\ &&
+{\frac {1}{2}}\int d^\nu \vec{r}_1 d^\nu \vec{r}_2 \int {\frac
{d\Omega}{\sqrt{2\pi}}}{\frac {d\omega_1}{\sqrt{2\pi}}} {\frac
{d\omega_2}{\sqrt{2\pi}}}{\tilde {\cal O}}(\Omega){\tilde
V}(\omega_1) {\tilde V}(\omega_2)
 \nonumber \\ &&  \hspace*{5cm}\times
e^{-{\frac {1}{2}}(\Omega^2+\omega_1^2+\omega_2^2) {\cal
U}-\omega_1\Omega f^{-1}_{\vec{r}_0 \vec{r}_1}-\omega_2\Omega
f^{-1}_{\vec{r}_0 \vec{r}_2}-\omega_1\omega_2 f^{-1}_{\vec{r}_1
\vec{r}_2}}
 \nonumber \\ &&
-{\frac {\mu^4}{8}}{\bar {\cal O}}^{(0)}\int d^\nu \vec{r}_1 d^\nu
\vec{r}_2 [2(f^{-1}_{\vec{r}_1 \vec{r}_2})^2+(f^{-1}_{\vec{r}_1
\vec{r}_1})^2]
 \nonumber \\ &&
+{\frac {\mu^2}{2}}{\bar {\cal O}}^{(0)}\int d^\nu \vec{r}_1 d^\nu
\vec{r}_2 \int {\frac {d\omega}{\sqrt{2\pi}}} {\tilde
V}(\omega)[f^{-1}_{\vec{r}_1
\vec{r}_1}+(i\omega)^2(f^{-1}_{\vec{r}_1 \vec{r}_2})^2]e^{-{\frac
{1}{2}}\omega^2 {\cal U}}
 \nonumber \\ &&
-{\frac {1}{2}}{\bar {\cal O}}^{(0)}\int d^\nu \vec{r}_1 d^\nu
\vec{r}_2 \int {\frac {d\omega_1}{\sqrt{2\pi}}} {\frac
{d\omega_2}{\sqrt{2\pi}}}{\tilde V}(\omega_1) {\tilde
V}(\omega_2)e^{-{\frac {1}{2}}(\omega_1^2+\omega_2^2) {\cal
U}-\omega_1\omega_2 f^{-1}_{\vec{r}_1 \vec{r}_2}}
 \nonumber \\ &&
-{\frac {\mu^2}{2}}{\bar {\cal O}}^{(1)}\int d^\nu \vec{r}_1
f^{-1}_{\vec{r}_1 \vec{r}_1}+{\bar {\cal O}}^{(1)}\int d^\nu
\vec{r}_1 \int {\frac {d\omega}{\sqrt{2\pi}}}{\tilde
V}(\omega)e^{-{\frac {1}{2}}\omega^2 {\cal U}}
 \nonumber \\
 &=&{\frac {\mu^4}{2}}\int d^\nu \vec{r}_1 d^\nu \vec{r}_2 f^{-1}_{\vec{r}_0
\vec{r}_1}f^{-1}_{\vec{r}_0 \vec{r}_2} f^{-1}_{\vec{r}_1
\vec{r}_2}\int^\infty_{-\infty}{\frac
{d\alpha}{\sqrt{2\pi}}}e^{-{\frac {\alpha^2}{2}}}{\cal
O}^{(2)}(\alpha\sqrt{\cal U})
 \nonumber \\ &&
+{\frac {\mu^4}{8}}\int d^\nu \vec{r}_1 d^\nu \vec{r}_2
(f^{-1}_{\vec{r}_0 \vec{r}_1})^2(f^{-1}_{\vec{r}_0 \vec{r}_2})^2
\int^\infty_{-\infty}{\frac {d\alpha}{\sqrt{2\pi}}}e^{-{\frac
{\alpha^2}{2}}}{\cal O}^{(4)}(\alpha\sqrt{\cal U})
 \nonumber \\ &&
-{\frac {\mu^2}{2}}\int d^\nu \vec{r}_1 d^\nu \vec{r}_2
(f^{-1}_{\vec{r}_0 \vec{r}_1})^2 \int^\infty_{-\infty}{\frac
{d\alpha d\gamma}{\sqrt{4\pi^2}}}e^{-{\frac {\alpha^2}{2}}-{\frac
{\gamma^2}{2}}} V(\gamma\sqrt{\cal U})
 \nonumber  \\ && \hspace*{5cm}
\times \Bigl[{\cal O}^{(2)}(\alpha\sqrt{{\cal W}_{02}}+
\gamma{\frac {f^{-1}_{\vec{r}_0 \vec{r}_2}}{\sqrt{{\cal U}}}})-
{\cal O}^{(2)}(\alpha\sqrt{{\cal U}})\Bigl]
 \nonumber  \\ &&
-{\frac {\mu^2}{2}}\int d^\nu \vec{r}_1 d^\nu \vec{r}_2
(f^{-1}_{\vec{r}_1 \vec{r}_2})^2 \int^\infty_{-\infty}{\frac
{d\alpha d\gamma}{\sqrt{4\pi^2}}}e^{-{\frac {\alpha^2}{2}}-{\frac
{\gamma^2}{2}}} V^{(2)}(\gamma\sqrt{\cal U})
 \nonumber  \\ && \hspace*{5cm}
 \times \Bigl[{\cal O}(\alpha\sqrt{{\cal W}_{02}}+ \gamma{\frac
{f^{-1}_{\vec{r}_0 \vec{r}_2}}{\sqrt{{\cal U}}}})-{\cal
O}(\alpha\sqrt{\cal U})\Bigl]
 \nonumber  \\ &&
- \mu^2\int d^\nu \vec{r}_1 d^\nu \vec{r}_2 f^{-1}_{\vec{r}_0
\vec{r}_1}f^{-1}_{\vec{r}_1 \vec{r}_2} \int^\infty_{-\infty}{\frac
{d\alpha d\gamma}{\sqrt{4\pi^2}}}e^{-{\frac {\alpha^2}{2}}-{\frac
{\gamma^2}{2}}} V^{(1)}(\gamma\sqrt{\cal U}) {\cal
O}^{(1)}(\alpha\sqrt{{\cal W}_{02}}+ \gamma{\frac
{f^{-1}_{\vec{r}_0 \vec{r}_2}}{\sqrt{{\cal U}}}})
 \nonumber  \\ &&
- {\frac {1}{2}}\int d^\nu \vec{r}_1 d^\nu \vec{r}_2
\int^\infty_{-\infty}{\frac {d\alpha d\gamma
d\lambda}{\sqrt{8\pi^3}}}e^{-{\frac {\alpha^2}{2}}-{\frac
{\gamma^2}{2}}-{\frac {\lambda^2}{2}}} V(\lambda\sqrt{\cal U})
V(\gamma\sqrt{{\cal W}_{12}}+\lambda {\frac {f^{-1}_{\vec{r}_1
\vec{r}_2}}{\sqrt{\cal U}}})\Bigl[{\cal O}(\alpha\sqrt{{\cal U}})
 \nonumber \\ && \hspace*{1.9cm}
-{\cal O}(\alpha\sqrt{{\cal W}_{02}-{\frac {({\cal
U}f^{-1}_{\vec{r}_0 \vec{r}_1}-f^{-1}_{\vec{r}_0
\vec{r}_2}f^{-1}_{\vec{r}_1 \vec{r}_2})^2}{{\cal U}^2{\cal
W}_{12}}}}+\gamma{\frac {{\cal U}f^{-1}_{\vec{r}_0
\vec{r}_1}-f^{-1}_{\vec{r}_0 \vec{r}_2}f^{-1}_{\vec{r}_1
\vec{r}_2}}{{\cal U}\sqrt{{\cal W}_{12}}}}+\lambda{\frac
{f^{-1}_{\vec{r}_0 \vec{r}_2}}{\sqrt{{\cal U}}}})\Bigl]
 \nonumber \\ &&
  - \int d^\nu \vec{r}_1 d^\nu
\vec{r}_2\int^\infty_{-\infty}{\frac {d\alpha d\gamma
d\lambda}{\sqrt{8\pi^3}}}e^{-{\frac {\alpha^2}{2}}-{\frac
{\gamma^2}{2}}-{\frac {\lambda^2}{2}}} V(\gamma \sqrt{\cal
U})V(\lambda \sqrt{\cal U})
 \nonumber  \\ && \hspace*{7cm} \times
\Bigl[{\cal O}(\alpha\sqrt{{\cal W}_{01}}+\gamma {\frac
{f^{-1}_{\vec{r}_0 \vec{r}_1}}{\sqrt{\cal U}}})-{\cal
O}(\alpha\sqrt{\cal U})\Bigl] \; .
\end{eqnarray}
Notice that in Eqs.(20) and (25), those expressions in all the
parentheses which follow the symbols $V$, $V^{(n)}$, ${\cal O}$
and ${\cal O}^{(n)}$ are arguments of the functions ${\cal O}(z)$
and ${\cal O}^{(n)}(z)$, respectively. From Eqs.(16), (20) and
(25), we can write down truncated VP expressions of ${\bar {\cal
O}}$ at the first order, ${\bar {\cal O}}^{I}={\bar {\cal
O}}^{(0)}+ \epsilon {\bar {\cal O}}^{(1)}$, and at the second
order, ${\bar {\cal O}}^{II}={\bar {\cal O}}^{(0)}+ \epsilon {\bar
{\cal O}}^{(1)}+ \epsilon^2 {\bar {\cal O}}^{(2)}$.

It seems that Eqs.(20) and (25) (so ${\bar {\cal O}}^{I}$ and
${\bar {\cal O}}^{II})$ have space-volume divergences because both
${\cal O}(\alpha\sqrt{\cal U})$ and ${\cal
O}^{(2)}(\alpha\sqrt{\cal U})$ are independent of space
coordinates $\vec{r}_1$ and $\vec{r}_2$. However, one can have a
simple analysis and see that the divergences really cancel.
Formally, one can write down the following expansion form
$${\cal
O}(\alpha\sqrt{{\cal W}_{01}}+\gamma {\frac {f^{-1}_{\vec{r}_0
\vec{r}_1}}{\sqrt{\cal U}}})={\cal O}(\alpha\sqrt{\cal
U})+\sum_{n=1}^{\infty}a_n (f^{-1}_{\vec{r}_0 \vec{r}_1})^n \;,$$
where the coefficients $a_n$s are functions of $\alpha,\gamma$ and
${\cal U}$ but independent of the coordinate $\vec{r}_1$, and
evidently, the similar expansion forms one can have for ${\cal
O}^{(2)}(\alpha\sqrt{{\cal W}_{02}}+ \gamma{\frac
{f^{-1}_{\vec{r}_0 \vec{r}_2}}{\sqrt{{\cal U}}}})$,${\cal
O}(\alpha\sqrt{{\cal W}_{02}}+ \gamma{\frac {f^{-1}_{\vec{r}_0
\vec{r}_2}}{\sqrt{{\cal U}}}})$,${\cal O}^{(1)}(\alpha\sqrt{{\cal
W}_{02}}+ \gamma{\frac {f^{-1}_{\vec{r}_0 \vec{r}_2}}{\sqrt{{\cal
U}}}})$ as well as $V(\gamma\sqrt{{\cal W}_{12}}+\lambda {\frac
{f^{-1}_{\vec{r}_1 \vec{r}_2}}{\sqrt{\cal U}}})$. Besides, one can
also has
 \begin{eqnarray*}
&&{\cal O}(\alpha\sqrt{{\cal W}_{02}-{\frac {({\cal
U}f^{-1}_{\vec{r}_0 \vec{r}_1}-f^{-1}_{\vec{r}_0
\vec{r}_2}f^{-1}_{\vec{r}_1 \vec{r}_2})^2}{{\cal U}^2{\cal
W}_{12}}}}+\gamma{\frac {{\cal U}f^{-1}_{\vec{r}_0
\vec{r}_1}-f^{-1}_{\vec{r}_0 \vec{r}_2}f^{-1}_{\vec{r}_1
\vec{r}_2}}{{\cal U}\sqrt{{\cal W}_{12}}}}+\lambda{\frac
{f^{-1}_{\vec{r}_0 \vec{r}_2}}{\sqrt{{\cal U}}}})
 \nonumber \\ &
=&{\cal O}(\alpha\sqrt{\cal U})+\sum_{l,m,n=0}^{\infty}a_{l,m,n}
(f^{-1}_{\vec{r}_0 \vec{r}_1})^l (f^{-1}_{\vec{r}_0 \vec{r}_2})^m
(f^{-1}_{\vec{r}_1 \vec{r}_2})^n \;,
 \end{eqnarray*}
where $l,m,n$ in the summation are required not to take zero
simultaneously, and the coefficients $a_{l,m,n}$s are independent
of the coordinates $\vec{r}_1$ and $\vec{r}_2$. Thus, there exist
really no space-volume divergences in Eqs.(20) and (25),
respectively. Consequently, both ${\bar {\cal O}}^{I}$ and ${\bar
{\cal O}}^{II})$ have no space-volume divergences.

Furthermore, substituting the above formal expansion forms into
Eqs.(20) and (25), one will find that in Eqs.(20) and (25), all
the integrals over $\vec{r}_1$ and $\vec{r}_2$ are involved only
in the following types
\begin{eqnarray*}
I^{(n)}&\equiv&\int d^\nu \vec{r}_1 (f^{-1}_{\vec{r}_0
\vec{r}_1})^n\nonumber \\
&=&\int d^\nu \vec{p}_1 d^\nu \vec{p}_2 \cdots d^\nu
\vec{p}_{n-1}[(2\pi)^{(n-1)\nu}(p_1^2+\mu^2)(p_2^2+\mu^2)
\cdots(p_{n-1}^2+\mu^2)((\sum_{i=1}^{n-1}\vec{p}_i)^2+\mu^2)]^{-1}
\;,
\end{eqnarray*}
\begin{eqnarray*}
I^{(l,m,n)}&=&\int d^\nu \vec{r}_1 d^\nu \vec{r}_2
(f^{-1}_{\vec{r}_0 \vec{r}_1})^l (f^{-1}_{\vec{r}_0 \vec{r}_2})^m
(f^{-1}_{\vec{r}_1 \vec{r}_1})^n =\int d^\nu \vec{p}_1 d^\nu
\vec{p}_2 \cdots d^\nu \vec{p}_{l} d^\nu \vec{p'}_1 d^\nu
\vec{p'}_2 \cdots d^\nu \vec{p'}_{m-1}
 \nonumber \\ &&\hspace*{1cm}\times
d^\nu \vec{p''}_1 d^\nu \vec{p''}_2 \cdots d^\nu \vec{p''}_{n-1}
[(2\pi)^{(l+m+n-2)\nu}\prod_{i=1}^{l}(p_i^2+\mu^2)\prod_{j=1}^{m-1}(|\vec{p'}_j|^2+\mu^2)
\end{eqnarray*}
\begin{eqnarray*}
 \hspace*{2cm}\times
\prod_{k=1}^{n-1}(|\vec{p''}_k|^2+\mu^2)((\sum_{i=1}^{l}\vec{p}_i+\sum_{j=1}^{m-1}\vec{p'}_j)^2+\mu^2)
((\sum_{i=1}^{l}\vec{p}_i-\sum_{k=1}^{n-1}\vec{p''}_k)^2+\mu^2)]^{-1}
 \end{eqnarray*}
and products of $I^{(n)}$s. In the case of $\nu=2$, ${\cal U}$ is
finite, $I^{(l,m,n)}$s as well as $I^{(n)}$ with any $l,m$ and $n$
are finite, and accordingly Eqs.(16), (20) and (25) are really
full of no explicit divergences. Since both ${\bar {\cal O}}^{I}$
and ${\bar {\cal O}}^{II}$ are finite when $\nu=2$, we can take
$\epsilon=1$ and perform the other key step of the VP scheme to
termine the parameter $\mu$ with the PMS. For the case of $\nu=2$,
according to the PMS, rendering ${\frac {\partial {\bar {\cal
O}}^{I}}{\partial \mu^2}}$ vanishing, one can solve it for
$\mu=\mu^I$ which is the value of $\mu$ at the first order, and
rendering ${\frac {\partial {\bar {\cal O}}^{II}}{\partial
\mu^2}}$ vanishing, one can solve it for $\mu=\mu^{II}$ which is
the value of $\mu$ up to the second order (in case the first
derivative condition couldn't produce meaningful root, the second
derivative condition ${\frac {\partial^2 {\bar {\cal
O}}^{II}}{(\partial \mu^2)^2}}=0$ will be used). Substituting
$\mu^I$ into ${\bar {\cal O}}^{I}$ gives the approximate result of
${\bar {\cal O}}$ up to the first order in the VP scheme, and
substituting $\mu^{II}$ into ${\bar {\cal O}}^{II}$ will give the
approximate result of ${\bar {\cal O}}$ up to the second order in
the VP scheme. Ref.~\cite{27} gave an example for doing so.

Presumably, although there are no explicit divergences in the case
of $\nu=2$, for non-polynomial potential $V(\phi_x)$ and/or
non-polynomial-type local fields ${\cal O}(\phi_{\vec{r}_0})$,
there possibly exist some terms which is not convergent for some
range of model parameters in the approximate results of ${\bar
{\cal O}}$. Furthermore, for $\nu=3$, although ${\cal U}$ is still
finite, but $I^{(n)}$ with any $n>2$ and $I^{(l,m,n)}$s with
$l+m+n>5$ are no longer finite, and for $\nu=4$, both ${\cal U}$
and $I^{(n)}$ with any $n>1$ and $I^{(l,m,n)}$s with $l+m+n>3$ are
divergent. For all the divergent cases, one have to appeal to a
further renormalization procedure before determining $\mu$ with
the PMS, as we have stated in last subsection. In the present
paper, we do not discuss it concretely.

In the same way, one can consider higher-order cases. Here we do
not continue it. In this section, we have developed a VP scheme of
calculating VEVs in QFT, and obtained the truncated VP expressions
for ${\bar {\cal O}}$ at the second order, Eqs.(16), (20) and
(25). The right hand sides of them are mostly Gaussian transforms
of the functions ${\cal O}(z)$, $V(z)$ and their derivatives
\cite{32}, and Eqs.(16), (20) and (25) are the aforementioned
Gaussian smearing formulae. According to these formulae, one can
easily obtain truncated VP expressions for VEVs of a local field
in a field theory by finishing only ordinary integrations.

\section{sG Field Theory: Truncations from the VP Series}
\label{3}

This section will derive the first three terms in Eq.(15) for
$G_a$ in three ways.

\subsection{Direct Performing the VP Expansion}
\label{diret}

We consider the $\nu$-dimensional Euclidean sG field theory with
the following Lagrangian density
\begin{equation}
{\cal L}_{sG}= {\frac {1}{2}}\nabla_{\vec{r}}\phi_{\vec{r}}
\nabla_{\vec{r}} \phi_{\vec{r}} - 2
\Omega\cos(\sqrt{8\pi}\beta\phi_{\vec{r}})\;.
\end{equation}
The Lagrangian density Eq.(26) is nothing but Eq.(5) in
Ref.~\cite{4} if one makes the transform $\phi\to {\frac
{\phi}{\sqrt{8\pi}}}$ (hereafter, we will use
$e^{i\sqrt{8\pi}a\phi(\vec{r}_0)}$ instead of
$e^{ia\phi(\vec{r}_0)}$ as the form of the local exponential field
and consequently the parameter $a$ and the coupling $\beta$ in
this paper are identical to those in Ref.~\cite{4}, respectively).
If taking $\sqrt{8\pi}\beta\to \beta$ and $2\Omega=m^2/\beta^2$
and adding the term $m^2/\beta^2$ in the Lagrangian density, one
will get the Euclidean version of the sG Lagrangian density which
discussed in Ref.~\cite{18}. Besides, taking the substitution
$\beta=i\gamma$ and $\Omega\to -\Delta$, Eq.(26) describes the
Euclidean shG field theory \cite{5,33}, and so the resultant
expressions in this section can easily be used to give the
corresponding expressions for the shG field theory.

In Eq.(26), $\beta$ is the coupling parameter with the dimension
[length$]^{(D-1)/2}$ and $\Omega$ is another parameter with the
dimension [length$]^{-D-1}$ in natural unit system. It is always
viable to have $\beta\ge 0$ without loss of generality. Obviously,
the classical potential $V(\phi_{\vec{r}})=-2
\Omega\cos(\sqrt{8\pi}\beta\phi_{\vec{r}})$ is invariant under the
transform $\phi\to \phi+{\frac {2\pi n}{\sqrt{8\pi}\beta}}$ with
any integer $n$, and so the classical vacua are infinitely
degenerate. So do the quantum vacua, as was shown, for example, by
the beyond-Gaussian effective potential for two-dimensional case
\cite{18}. Here, we choose to consider the vacuum with the
expectation value of the sG field operator $\phi_{\vec{r}}$
vanishing.

According to the definition Eq.(8), the VEV of the local
exponential field $e^{i\sqrt{8\pi}a\phi(\vec{r}_0)}$ in the sG
field theory is defined as follows
\begin{equation}
G_a\equiv <e^{i \sqrt{8\pi}a \phi(0)}>\equiv {\frac {\int {\cal
D}\phi \exp\{i \sqrt{8\pi}a \phi(0)\} \exp\{-\int d^\nu \vec{r}{\cal
L}_{sG}\}} {\int {\cal D}\phi \exp\{-\int d^\nu \vec{r}{\cal
L}_{sG}\}}}  \;.
\end{equation}
For simplicity, the exponential field in Eq.(27) is taken at
$\vec{r}_0=0$. It is evident that the numerator and denominator in
the right hand side of Eq.(27) can be easily got from the sG
generating functional
\begin{equation}
Z_{sG}[J]=\int {\cal D}\phi \exp\{-\int d^\nu \vec{r}\; [{\cal
L}_{sG}-J_{\vec{r}}\phi_{\vec{r}}\; ]\}  \; ,
\end{equation}
by taking $J_{\vec{r}}=i \sqrt{8\pi}a \delta(\vec{r})$ and $J=0$,
respectively.

Doing as was done for getting to Eqs.(14) and (15) in last
section, one can obtain a similar expansion for $G_a$. In the
present case, all the things are same as in subsection A of
Sect.II, except for, now, ${\cal N}_{\cal M}[{\cal
O}(\phi_{\vec{r}_0})]=\exp\{4\pi a^2 I_{(1)}({\cal
M}^2)\}e^{i\sqrt{8\pi}a\phi(0)}$ and ${\cal N}_{\cal
M}[V(\phi_{\vec{r}})]=-2\Omega\exp\{4\pi \beta^2 I_{(1)}({\cal
M}^2)\}cos{\sqrt{8\pi}\beta\phi_{\vec{r}}}$. Consequently, one can
have
\begin{equation}
G_a = e^{4\pi a^2 I_{(1)}({\cal M}^2)}{\frac
{\bigl[\sum_{n=0}^\infty \epsilon^n {\frac {(-1)^{n}}{n!}} \int
\prod_{k=1}^n d^\nu \vec{r}_k {\cal H}^{sG}_I({\frac
{\delta}{\delta J_{\vec{r}_k}}},\mu)\exp\{{\frac
{1}{2}}Jf^{-1}J\}\bigl]_{J_{\vec{r}}=i \sqrt{8\pi}a
\delta(\vec{r})}}{\bigl[\sum_{n=0}^\infty \epsilon^n {\frac
{(-1)^{n}}{n!}} \int \prod_{k=1}^n d^\nu \vec{r}_k {\cal
H}^{sG}_I({\frac {\delta}{\delta J_{\vec{r}_k}}},\mu)
     \exp\{{\frac {1}{2}}Jf^{-1}J\}\bigl]_{J=0}}}\Biggl|_{\epsilon=1}
\end{equation}
with
\begin{equation}
{\cal H}^{sG}_I(\phi_{\vec{r}},\mu)=-{\frac
{1}{2}}\mu^2\phi_{\vec{r}}^2-2\Omega\cos(\sqrt{8\pi}\beta\phi_{\vec{r}})\exp\{4\pi\beta^2
I_{(1)}({\cal M}^2)\}
\end{equation}
and the series consisting of powers of $\epsilon$
\begin{equation}
G_a=[G_a^{(0)}+\epsilon^1 G_a^{(1)}+\epsilon^2 G_a^{(2)}+\cdots
+\epsilon^n G_a^{(n)} +\cdots]_{\epsilon=1} =\Bigl[
\sum_{n=0}^\infty \epsilon^n G_a^{(n)}\Bigl]_{\epsilon=1} \;.
\end{equation}
Next, we will derive the first three terms in Eq.(31).

At the zeroth order, Eqs.(29) and (31) lead to
\begin{eqnarray}
G_a^{(0)}&=&\exp\{4\pi a^2 I_{(1)}({\cal M}^2)\}\Bigl[\exp\{{\frac
{1}{2}}Jf^{-1}J\}\Bigl]_{J_{\vec{r}}=i \sqrt{8\pi}a
\delta(\vec{r})}\nonumber  \\ &=&e^{-4\pi a^2 {\cal U}}  \;.
\end{eqnarray}

At the first order, the coefficient $G_a^{(1)}$ is as follows
\begin{eqnarray}
G_a^{(1)}&=&-\exp\{4\pi a^2 I_{(1)}({\cal M}^2)\}\Bigl[\int d^\nu
\vec{r}_1 {\cal H}^{sG}_I({\frac {\delta}{\delta
J_{\vec{r}_1}}},\mu) \exp\{{\frac
{1}{2}}Jf^{-1}J\}\Bigl]_{J_{\vec{r}}=i \sqrt{8\pi}a
\delta(\vec{r})} \nonumber
\\ &&+G_a^{(0)}\Bigl[\int d^\nu \vec{r}_1 {\cal H}^{sG}_I({\frac {\delta}{\delta
J_{\vec{r}_1}}},\mu)\exp\{{\frac
{1}{2}}Jf^{-1}J\}\Bigl]_{J_{\vec{r}}=0} \;.
\end{eqnarray}
To calculate $G_a^{(1)}$, one can first have
\begin{eqnarray}
&&{\cal H}^{sG}_I({\frac {\delta}{\delta J_{\vec{r}_1}}},\mu)
\exp\{{\frac {1}{2}}Jf^{-1}J\}
 \nonumber  \\
&=& -{\frac {1}{2}}\mu^2{\frac {\delta^2}{\delta
J_{\vec{r}_1}^2}}\exp\{{\frac
{1}{2}}Jf^{-1}J\}-\Omega\exp\{4\pi\beta^2 I_{(1)}({\cal M}^2)\}
\Bigl[\exp\{{\frac {1}{2}}Jf^{-1}J\}\bigl|_{J_{\vec{r}}\to
J_{\vec{r}}+i \sqrt{8\pi}\beta \delta(\vec{r}-\vec{r}_1)}
 \nonumber  \\
&& \hspace*{6cm} +\exp\{{\frac
{1}{2}}Jf^{-1}J\}\bigl|_{J_{\vec{r}}\to J_{\vec{r}}-i
\sqrt{8\pi}\beta \delta(\vec{r}-\vec{r}_1)}\Bigl] \;.
\end{eqnarray}
In writing down Eq.(34), we first returned the cosine part of its
left hand side into the original functional form, and then used
exponential form of cosine function and the result Eq.(6).
Substituting Eq.(34) together with Eq.(18) into (33), one obtains
\begin{eqnarray}
G_a^{(1)}&=&-4\pi \mu^2 a^2 e^{-4\pi a^2 {\cal U}}\int d^\nu
\vec{r}_1 (f^{-1}_{0 \vec{r}_1})^2 \nonumber  \\ && +2 \Omega
e^{-4\pi (a^2+\beta^2) {\cal U}}\sum_{n=1}^\infty {\frac {(8\pi a
\beta)^{2 n}}{(2 n)!}}\int d^\nu \vec{r}_1 (f^{-1}_{0
\vec{r}_1})^{2n}  \;.
\end{eqnarray}

Finally, we consider the second order. From Eqs.(29) and (31) ,
one can write down the second-order coefficient in Eq.(31) as
\begin{eqnarray}
G_a^{(2)}&=&{\frac {1}{2!}}e^{4\pi a^2 I_{(1)}({\cal
M}^2)}\Bigl[\int d^\nu \vec{r}_1 d^\nu \vec{r}_2 {\cal
H}^{sG}_I({\frac {\delta}{\delta J_{\vec{r}_1}}},\mu){\cal
H}^{sG}_I({\frac {\delta}{\delta J_{\vec{r}_2}}},\mu)\exp\{{\frac
{1}{2}}Jf^{-1}J\}
\Bigl]_{J_{\vec{r}}=i \sqrt{8\pi}a \delta(\vec{r})}\nonumber \\
&&-{\frac {1}{2!}}G_a^{(0)}\Bigl[\int d^\nu \vec{r}_1 d^\nu
\vec{r}_2 {\cal H}^{sG}_I({\frac {\delta}{\delta
J_{\vec{r}_1}}},\mu){\cal H}^{sG}_I({\frac {\delta}{\delta
J_{\vec{r}_2}}},\mu)\exp\{{\frac
{1}{2}}Jf^{-1}J\} \Bigl]_{J_{\vec{r}}=0} \nonumber \\
&&+G_a^{(1)}\Bigl[\int d^\nu \vec{r}_1 {\cal H}^{sG}_I({\frac
{\delta}{\delta J_{\vec{r}_1}}},\mu) \exp\{{\frac
{1}{2}}Jf^{-1}J\}\Bigl]_{J_{\vec{r}}=0}\;.
\end{eqnarray}
Doing as was done at the first order, one can first have
\begin{eqnarray}
&&{\cal H}_I^{sG}({\frac {\delta}{\delta J_{\vec{r}_1}}},\mu){\cal
H}_I^{sG}({\frac {\delta}{\delta J_{\vec{r}_2}}},\mu) \exp\{{\frac
{1}{2}}Jf^{-1}J\} \nonumber \\ &=&{\frac {1}{4}}\mu^4 {\frac
{\delta^2}{\delta J_{\vec{r}_1}^2}}{\frac {\delta^2}{\delta
J_{\vec{r}_2}^2}}\exp\{{\frac {1}{2}}Jf^{-1}J\}
 \nonumber \\ &&+\mu^2\Omega \exp\{4\pi\beta^2 I_{(1)}({\cal
 M}^2)\} {\frac {\delta^2}{\delta
J_{\vec{r}_1}^2}}\biggl[\exp\{{\frac
{1}{2}}Jf^{-1}J\}\bigl|_{J_{\vec{r}}\to J_{\vec{r}}+i
\sqrt{8\pi}\beta \delta(\vec{r}-\vec{r}_2)} \nonumber
\\ && \hspace*{5cm} +\exp\{{\frac {1}{2}}Jf^{-1}J\}\bigl|_{J_{\vec{r}}\to
J_{\vec{r}}-i
\sqrt{8\pi}\beta \delta(\vec{r}-\vec{r}_2)}\biggl] \nonumber \\
&&+\Omega^2 \exp\{8\pi\beta^2 I_{(1)}({\cal
 M}^2)\} \biggl[\exp\{{\frac
{1}{2}}Jf^{-1}J\}\bigl|_{J_{\vec{r}}\to J_{\vec{r}}+i
\sqrt{8\pi}\beta \delta(\vec{r}-\vec{r}_1)+i \sqrt{8\pi}\beta
\delta(\vec{r}-\vec{r}_2)} \nonumber  \\ && \hspace*{5cm}+2
\exp\{{\frac {1}{2}}Jf^{-1}J\}\bigl|_{J_{\vec{r}}\to J_{\vec{r}}+i
\sqrt{8\pi}\beta \delta(\vec{r}-\vec{r}_1)-i \sqrt{8\pi}\beta
\delta(\vec{r}-\vec{r}_2)} \nonumber  \\ &&
\hspace*{5cm}+\exp\{{\frac {1}{2}}Jf^{-1}J\}\bigl|_{J_{\vec{r}}\to
J_{\vec{r}}-i \sqrt{8\pi}\beta \delta(\vec{r}-\vec{r}_1)-i
\sqrt{8\pi}\beta \delta(\vec{r}-\vec{r}_2)}\biggl]  \;.
\end{eqnarray}
Substituting Eq.(22) into Eq.(37) leads to
\begin{eqnarray*}
&&{\cal H}_I^{sG}({\frac {\delta}{\delta J_{\vec{r}_1}}},\mu){\cal
H}_I^{sG}({\frac {\delta}{\delta J_{\vec{r}_2}}},\mu) \exp\{{\frac
{1}{2}}Jf^{-1}J\}
 \nonumber \\ &=& {\frac {1}{4}}\mu^4
\bigl\{(f^{-1}_{\vec{r_1}\vec{r_1}})^2+2(f^{-1}_{\vec{r_1}\vec{r_2}})^2+2
f^{-1}_{\vec{r_1}\vec{r_1}} (Jf^{-1})_1^2
 \nonumber \\
&&\hspace*{2cm}+4 f^{-1}_{\vec{r_1}\vec{r_2}} (Jf^{-1})_1
(Jf^{-1})_2 +(Jf^{-1})_1^2 (Jf^{-1})_2^2\}\exp\{{\frac
{1}{2}}Jf^{-1}J\bigl\}
 \nonumber \\ &&+\mu^2\Omega \exp\{4\pi\beta^2 I_{(1)}({\cal
 M}^2)\}\bigl\{2[f^{-1}_{\vec{r_1}\vec{r_1}}+(Jf^{-1})_1^2+8\pi (i\beta)^2 (f^{-1}_{\vec{r_1}\vec{r_2}})^2]
 \cos[\sqrt{8\pi}\beta(Jf^{-1})_2]
 \nonumber \\ && \hspace*{2cm}- 4 \sqrt{8\pi}\beta f^{-1}_{\vec{r_1}\vec{r_2}}(Jf^{-1})_1
 \sin[\sqrt{8\pi} \beta(Jf^{-1})_2]\}
 \exp\{{\frac {1}{2}}Jf^{-1}J+ 4\pi (i\beta)^2 (f^{-1}_{\vec{r_2}\vec{r_2}})^2\bigl\}
\end{eqnarray*}
\begin{eqnarray}
 && +\Omega^2 \exp\{8\pi\beta^2 I_{(1)}({\cal
 M}^2)\}
 \nonumber \\ && \hspace*{0.5cm}\times \bigl\{ 2 \exp\{{\frac {1}{2}}Jf^{-1}J+ i\sqrt{8\pi}\beta [(Jf^{-1})_1-(Jf^{-1})_2]
 +8\pi(i\beta)^2[f^{-1}_{\vec{r_1}\vec{r_1}}-f^{-1}_{\vec{r_1}\vec{r_2}}]\} \nonumber \\ &&
 \hspace*{1cm}+ 2 \cos[\sqrt{8\pi}\beta((Jf^{-1})_1+(Jf^{-1})_2)] \exp\{{\frac {1}{2}}Jf^{-1}J
 + 8\pi(i\beta)^2
 [f^{-1}_{\vec{r_1}\vec{r_1}}+f^{-1}_{\vec{r_1}\vec{r_2}}]\}\bigl\}
 \; .
\end{eqnarray}
For simplicity, in writing down last equation, we have used the
the property that Eq.(36) is invariant under interchanging
$\vec{r}_1$ and $\vec{r}_2$. (By the way, for getting
Eqs.(13),(14) and (16) in Ref.~\cite{18}, we also used the similar
property, and in the right hand side of Eq.(16) in Ref.~\cite{18},
the first term should have an additional factor ``$\; {\frac
{1}{2}} \; e^{Jf^{-1}J/2}\;$" and the second term should have an
additional factor ``$\; {\frac {1}{2}} \;$ ".) Thus, substituting
Eqs.(32),(35) and (38) into Eq.(36), one can eventually obtain
\begin{eqnarray}
G_a^{(2)}&=& - 4\pi a^2\mu^4e^{-4\pi a^2 {\cal U}}\int d^\nu
\vec{r}_1 d^2 \vec{r}_2 f^{-1}_{\vec{r}_1 \vec{r}_2} f^{-1}_{0
\vec{r}_1}f^{-1}_{0 \vec{r}_2} +8\pi^2 \mu^4 a^4 e^{-4\pi a^2
{\cal U}}\Bigl[\int d^\nu \vec{r}_1 (f^{-1}_{0
\vec{r}_1})^2\Bigl]^2
 \nonumber\\ &&
-8\pi \mu^2\Omega a^2e^{-4\pi (a^2+\beta^2) {\cal U}}\int d^\nu
\vec{r}_1 (f^{-1}_{0 \vec{r}_1})^2\int d^\nu \vec{r}_2 [\cosh(8\pi
a \beta f^{-1}_{0 \vec{r}_2})-1]
 \nonumber \\ &&
-8\pi \mu^2\Omega \beta^2e^{-4\pi (a^2+\beta^2) {\cal U}}\int
d^\nu \vec{r}_1 d^\nu \vec{r}_2 (f^{-1}_{\vec{r}_1 \vec{r}_2})^2
[\cosh(8\pi a \beta f^{-1}_{0 \vec{r}_2})-1]
 \nonumber \\ &&
+ 16\pi \mu^2 \Omega a \beta e^{-4\pi (a^2+\beta^2) {\cal U}} \int
d^\nu \vec{r}_1 d^\nu \vec{r}_2 f^{-1}_{\vec{r}_1 \vec{r}_2}
f^{-1}_{0 \vec{r}_1}\sinh(8\pi a \beta f^{-1}_{0 \vec{r}_2})
 \nonumber \\&&
+2\Omega^2 e^{-4\pi (a^2+2\beta^2) {\cal U}} \Bigl\{\int d^\nu
\vec{r}_1 [\cosh(8\pi a \beta f^{-1}_{0 \vec{r}_1})-1]\Bigl\}^2
 \nonumber \\ &&
+ \Omega^2 e^{-4\pi (a^2+2\beta^2) {\cal U}} \int d^\nu \vec{r}_1
d^2 \vec{r}_2 [e^{-8\pi\beta^2 f^{-1}_{\vec{r}_1 \vec{r}_2}}-1]
 [\cosh(8\pi a \beta (f^{-1}_{0
\vec{r}_1}+f^{-1}_{0 \vec{r}_2}))-1]
 \nonumber \\&&
+\Omega^2 e^{-4\pi (a^2+2\beta^2) {\cal U}}\int d^\nu \vec{r}_1
d^\nu \vec{r}_2 [e^{8\pi\beta^2 f^{-1}_{\vec{r}_1
\vec{r}_2}}-1][e^{-8\pi a \beta (f^{-1}_{0 \vec{r}_1}-f^{-1}_{0
\vec{r}_2})}-1]\;.
\end{eqnarray}

From Eqs.(32), (35) and (39), one can easily write down
$G_a^{I}=G_a^{(0)}+ \epsilon G_a^{(1)}$ and $G_a^{II}=G_a^{(0)}+
\epsilon G_a^{(1)}+ \epsilon^2 G_a^{(2)}$, which are the truncated
expressions from the VP expansion series, Eq.(31), at the first
order and the second order, respectively. It is evident that
Eqs.(32), (35) and (39) with $\nu=2$ have not any explicit
divergences, providing an example for our analysis in last
section. Using the result ${\cal U}=-{\frac {1}{4 \pi}}\ln({\frac
{\mu^2}{{\cal M}^2}})$, Eqs.(32), (35) and (39) with $\nu=2$ can
lead to Eq.(12) in Ref.~\cite{27}.

\subsection{Using Gaussian Smearing Formulae}
\label{gsf}

This subsection will substitute the concrete expressions of the sG
potential and the local exponential field into Eqs.(16), (20) and
(25), respectively, and finish those Gaussian transforms to
recover Eqs.(32), (35) and (39).

First, from Eq.(16), $G_a^{(0)}$ can be written as
\begin{equation}
G_a^{(0)}=\int_{-\infty}^{\infty} {\frac
{d\alpha}{\sqrt{2\pi}}}e^{-{\frac {\alpha^2}{2}}}e^{i \sqrt{8\pi} a
 \alpha \sqrt{\cal U}} \;.
\end{equation}
Employing the Gaussian integral formula between Eqs.(16) and (17),
one can easily check that last equation really gives the result
Eq.(32).

Second, according to Eq.(20), $G_a^{(1)}$ takes the following form
\begin{eqnarray}
G_a^{(1)}&=&{\frac {1}{2}}\mu^2\int d^\nu \vec{r}_1 (f^{-1}_{0
\vec{r}_1})^2 \int^\infty_{-\infty}{\frac
{d\alpha}{\sqrt{2\pi}}}e^{-{\frac {\alpha^2}{2}}}(-8\pi a^2)e^{i
\sqrt{8\pi} a \alpha\sqrt{\cal U}} \nonumber
 \\ && + \int d^\nu \vec{r}_1 \int^\infty_{-\infty}{\frac
{d\gamma}{\sqrt{2\pi}}}e^{-{\frac {\gamma^2}{2}}} 2\Omega
\cos(\sqrt{8\pi} \beta\gamma
\sqrt{\cal U}) \nonumber  \\
&&  \hspace*{1cm}\times \int^\infty_{-\infty}{\frac
{d\alpha}{\sqrt{2\pi}}}e^{-{\frac {\alpha^2}{2}}}\Bigl[e^{i
\sqrt{8\pi} a (\alpha\sqrt{{\cal W}_{01}}+\gamma {\frac {f^{-1}_{0
\vec{r}_1}}{\sqrt{\cal U}}})}-e^{i \sqrt{8\pi} a \alpha\sqrt{\cal
U}}\Bigl] \;.
\end{eqnarray}
Finishing integrations over $\alpha$ and $\gamma$ in last equation
with the help of the Gaussian integral formula, one can have
\begin{equation}
G_a^{(1)}= -4\pi \mu^2 a^2 \int d^\nu \vec{r}_1 (f^{-1}_{0
\vec{r}_1})^2 e^{-4\pi a^2{\cal U}} +2\Omega \int d^\nu \vec{r}_1
e^{-4\pi (a^2+\beta^2){\cal U}}
 [\cosh(8\pi a \beta f^{-1}_{0 \vec{r}_1})-1]
 \;.
\end{equation}
Expanding $\cosh(8\pi a \beta f^{-1}_{0 \vec{r}_1})$ in Eq.(42) as
power series leads to nothing but Eq.(35).

Eq.(39) can also be obtained from the Gaussian smearing formulae.
Substituting $V(\phi_x)$ and ${\cal O}(\phi_{\vec {r}_0=0})$ for
the sG field theory into Eq.(25), one has
\begin{eqnarray*}
G_a^{(2)}&=&-4\pi a^2\mu^4\int d^\nu \vec{r}_1 d^\nu \vec{r}_2
f^{-1}_{\vec{r}_0 \vec{r}_1}f^{-1}_{\vec{r}_0 \vec{r}_2}
f^{-1}_{\vec{r}_1 \vec{r}_2}\int^\infty_{-\infty}{\frac
{d\alpha}{\sqrt{2\pi}}}e^{-{\frac {\alpha^2}{2}}}e^{i a
\sqrt{8\pi\cal U}\alpha}
 \nonumber \\ &&
+8\pi^2 a^4\mu^4\int d^\nu \vec{r}_1 d^\nu \vec{r}_2
(f^{-1}_{\vec{r}_0 \vec{r}_1})^2(f^{-1}_{\vec{r}_0 \vec{r}_2})^2
\int^\infty_{-\infty}{\frac {d\alpha}{\sqrt{2\pi}}}e^{-{\frac
{\alpha^2}{2}}}e^{i a \sqrt{8\pi\cal U}\alpha}
 \nonumber \\ &&
-4\pi a^2\mu^2\Omega\int d^\nu \vec{r}_1 d^\nu \vec{r}_2
(f^{-1}_{\vec{r}_0 \vec{r}_1})^2 \int^\infty_{-\infty}{\frac
{d\alpha}{\sqrt{2\pi}}}{\frac {d\gamma}{\sqrt{2\pi}}}e^{-{\frac
{\alpha^2}{2}}-{\frac {\gamma^2}{2}}}
 \nonumber  \\ && \hspace*{2cm}
\times \Bigl\{e^{i a \sqrt{8\pi{\cal W}_{02}}\alpha}\bigl[e^{i
\sqrt{8\pi}(a {\frac {f^{-1}_{\vec{r}_0 \vec{r}_2}}{\sqrt{{\cal
U}}}}+\beta\sqrt{{\cal U}})\gamma}+e^{i \sqrt{8\pi}(a {\frac
{f^{-1}_{\vec{r}_0 \vec{r}_2}}{\sqrt{{\cal U}}}}-\beta\sqrt{{\cal
U}})\gamma}\bigl]
 \nonumber  \\ && \hspace*{3cm}
-e^{i a \sqrt{8\pi{\cal U}}\alpha}\bigl[e^{i \beta\sqrt{8\pi{\cal
U}}\gamma}+e^{-i \beta\sqrt{8\pi{\cal U}}\gamma}\bigl]\Bigl\}
 \nonumber  \\ &&
-4\pi \beta^2\mu^2\Omega\int d^\nu \vec{r}_1 d^\nu \vec{r}_2
(f^{-1}_{\vec{r}_1 \vec{r}_2})^2 \int^\infty_{-\infty}{\frac
{d\alpha}{\sqrt{2\pi}}}{\frac {d\gamma}{\sqrt{2\pi}}}e^{-{\frac
{\alpha^2}{2}}-{\frac {\gamma^2}{2}}}
 \nonumber  \\ && \hspace*{2cm}
\times \Bigl\{e^{i a \sqrt{8\pi{\cal W}_{02}}\alpha}\bigl[e^{i
\sqrt{8\pi}(a {\frac {f^{-1}_{\vec{r}_0 \vec{r}_2}}{\sqrt{{\cal
U}}}}+\beta\sqrt{{\cal U}})\gamma}+e^{i \sqrt{8\pi}(a {\frac
{f^{-1}_{\vec{r}_0 \vec{r}_2}}{\sqrt{{\cal U}}}}-\beta\sqrt{{\cal
U}})\gamma}\bigl]
 \nonumber  \\ && \hspace*{3cm}
-e^{i a \sqrt{8\pi{\cal U}}\alpha}\bigl[e^{i \beta\sqrt{8\pi{\cal
U}}\gamma}+e^{-i \beta\sqrt{8\pi{\cal U}}\gamma}\bigl]\Bigl\}
 \nonumber  \\ &&
-8\pi a \beta \Omega\mu^2\int d^\nu \vec{r}_1 d^\nu \vec{r}_2
f^{-1}_{\vec{r}_0 \vec{r}_1}f^{-1}_{\vec{r}_1 \vec{r}_2}
\int^\infty_{-\infty}{\frac {d\alpha}{\sqrt{2\pi}}}{\frac
{d\gamma}{\sqrt{2\pi}}}e^{-{\frac {\alpha^2}{2}}-{\frac
{\gamma^2}{2}}}
\end{eqnarray*}
\begin{eqnarray}
 \hspace*{1cm}&&\hspace*{4cm}
\times e^{i a \sqrt{8\pi{\cal W}_{02}}\alpha}\bigl[e^{i
\sqrt{8\pi}(a {\frac {f^{-1}_{\vec{r}_0 \vec{r}_2}}{\sqrt{{\cal
U}}}}+\beta\sqrt{{\cal U}})\gamma}-e^{i \sqrt{8\pi}(a {\frac
{f^{-1}_{\vec{r}_0 \vec{r}_2}}{\sqrt{{\cal U}}}}-\beta\sqrt{{\cal
U}})\gamma}\bigl]
 \nonumber  \\ &&
+ {\frac {1}{2}}\Omega^2\int d^\nu \vec{r}_1 d^\nu \vec{r}_2
\int^\infty_{-\infty}{\frac {d\alpha}{\sqrt{2\pi}}}{\frac
{d\gamma}{\sqrt{2\pi}}}{\frac {d\lambda}{\sqrt{2\pi}}}e^{-{\frac
{\alpha^2}{2}}-{\frac {\gamma^2}{2}}-{\frac {\lambda^2}{2}}}
 \nonumber \\ && \hspace*{0.2cm}
\times \biggl\{e^{i a \sqrt{8\pi({\cal W}_{02}-{\frac {({\cal
U}f^{-1}_{\vec{r}_0 \vec{r}_1}-f^{-1}_{\vec{r}_0
\vec{r}_2}f^{-1}_{\vec{r}_1 \vec{r}_2})^2}{{\cal U}^2{\cal
W}_{12}}})}\alpha}\Bigl\{e^{i \sqrt{8\pi}(\beta\sqrt{{\cal W}_{12}}+
a {\frac {{\cal U}f^{-1}_{\vec{r}_0 \vec{r}_1}-f^{-1}_{\vec{r}_0
\vec{r}_2}f^{-1}_{\vec{r}_1 \vec{r}_2}}{{\cal U}\sqrt{{\cal
W}_{12}}}})\gamma}
 \nonumber \\ && \hspace*{4cm}
\times \bigl[e^{i \sqrt{8\pi}[\beta(\sqrt{{\cal U}}+{\frac
{f^{-1}_{\vec{r}_1 \vec{r}_2}}{\sqrt{{\cal U}}}})+a {\frac
{f^{-1}_{\vec{r}_0 \vec{r}_2}}{\sqrt{{\cal U}}}}]\lambda}+e^{i
\sqrt{8\pi}[\beta(-\sqrt{{\cal U}}+{\frac {f^{-1}_{\vec{r}_1
\vec{r}_2}}{\sqrt{{\cal U}}}})+a {\frac {f^{-1}_{\vec{r}_0
\vec{r}_2}}{\sqrt{{\cal U}}}}]\lambda}\bigl]
 \nonumber\\ &&\hspace*{6cm}
+e^{i \sqrt{8\pi}(-\beta\sqrt{{\cal W}_{12}}+ a {\frac {{\cal
U}f^{-1}_{\vec{r}_0 \vec{r}_1}-f^{-1}_{\vec{r}_0
\vec{r}_2}f^{-1}_{\vec{r}_1 \vec{r}_2}}{{\cal U}\sqrt{{\cal
W}_{12}}}})\gamma}
  \nonumber \\ && \hspace*{4cm}
\times \bigl[e^{i \sqrt{8\pi}[\beta(\sqrt{{\cal U}}-{\frac
{f^{-1}_{\vec{r}_1 \vec{r}_2}}{\sqrt{{\cal U}}}})+a {\frac
{f^{-1}_{\vec{r}_0 \vec{r}_2}}{\sqrt{{\cal U}}}}]\lambda}+e^{i
\sqrt{8\pi}[-\beta(\sqrt{{\cal U}}+{\frac {f^{-1}_{\vec{r}_1
\vec{r}_2}}{\sqrt{{\cal U}}}})+a {\frac {f^{-1}_{\vec{r}_0
\vec{r}_2}}{\sqrt{{\cal U}}}}]\lambda}\bigl]\Bigl\}
 \nonumber \\ && \hspace*{1cm}
-e^{ia\sqrt{8\pi{\cal U}}\alpha}\Bigl\{e^{i \gamma \beta
\sqrt{8\pi{\cal W}_{12}}}\bigl[e^{i
\lambda\beta\sqrt{8\pi}(\sqrt{{\cal U}}+{\frac {f^{-1}_{\vec{r}_1
\vec{r}_2}}{\sqrt{{\cal U}}}})}+e^{i
\lambda\beta\sqrt{8\pi}(-\sqrt{{\cal U}}+{\frac {f^{-1}_{\vec{r}_1
\vec{r}_2}}{\sqrt{{\cal U}}}})}\bigl]
 \nonumber \\ && \hspace*{2cm}
+e^{-i \gamma \beta \sqrt{8\pi{\cal W}_{12}}}\bigl[e^{i
\lambda\beta\sqrt{8\pi}(\sqrt{{\cal U}}-{\frac {f^{-1}_{\vec{r}_1
\vec{r}_2}}{\sqrt{{\cal U}}}})}+e^{-i
\lambda\beta\sqrt{8\pi}(\sqrt{{\cal U}}+{\frac {f^{-1}_{\vec{r}_1
\vec{r}_2}}{\sqrt{{\cal U}}}})}\bigl]\Bigl\}\biggl\}
 \nonumber \\ &&
- \Omega^2\int d^\nu \vec{r}_1 d^\nu \vec{r}_2
\int^\infty_{-\infty}{\frac {d\alpha}{\sqrt{2\pi}}}{\frac
{d\gamma}{\sqrt{2\pi}}}{\frac {d\lambda}{\sqrt{2\pi}}}e^{-{\frac
{\alpha^2}{2}}-{\frac {\gamma^2}{2}}-{\frac {\lambda^2}{2}}}
 \nonumber \\ && \hspace*{0.2cm}
\times \Bigl\{e^{i a \sqrt{8\pi{\cal W}_{01}}\alpha}\bigl[e^{i
\lambda\sqrt{8\pi}(a {\frac {f^{-1}_{\vec{r}_0
\vec{r}_1}}{\sqrt{{\cal U}}}}+\beta\sqrt{{\cal U}})}+e^{i
\lambda\sqrt{8\pi}(a {\frac {f^{-1}_{\vec{r}_0
\vec{r}_1}}{\sqrt{{\cal U}}}}-\beta\sqrt{{\cal U}})}\bigl]
\times\bigl[e^{i \gamma\beta \sqrt{8\pi{\cal U}}}+e^{-i \gamma\beta
\sqrt{8\pi{\cal U}}}\bigl]
 \nonumber \\ &&
\hspace*{1cm}- e^{i a \sqrt{8\pi{\cal U}}\alpha}\bigl[e^{i
\gamma\beta \sqrt{8\pi{\cal U}}}+e^{-i \gamma\beta \sqrt{8\pi{\cal
U}}}\bigl]\bigl[e^{i \lambda\beta\sqrt{8\pi{\cal U}}}+e^{-i
\lambda\beta\sqrt{8\pi{\cal U}}}\bigl]\Bigl\} \;.
\end{eqnarray}
Carrying out all integrations over $\alpha$, $\gamma$ and
$\lambda$ in last equation with the aid of the Gaussian integral
formula, one can get to
\begin{eqnarray}
G_a^{(2)}&=&-4\pi a^2\mu^4e^{-4\pi a^2{\cal U}}\int d^\nu \vec{r}_1
d^\nu \vec{r}_2 f^{-1}_{\vec{r}_0 \vec{r}_1}f^{-1}_{\vec{r}_0
\vec{r}_2} f^{-1}_{\vec{r}_1 \vec{r}_2}
  \nonumber \\ &&
+8\pi^2 a^4\mu^4e^{-4\pi a^2{\cal U}}\int d^\nu \vec{r}_1 d^\nu
\vec{r}_2 (f^{-1}_{\vec{r}_0 \vec{r}_1})^2(f^{-1}_{\vec{r}_0
\vec{r}_2})^2
 \nonumber \\ &&
-8\pi a^2\mu^2\Omega e^{-4\pi(a^2+\beta^2){\cal U}}\int d^\nu
\vec{r}_1 d^\nu \vec{r}_2 (f^{-1}_{\vec{r}_0 \vec{r}_1})^2
[\cosh(8\pi a \beta f^{-1}_{\vec{r}_0 \vec{r}_2})-1]
 \nonumber \\ &&
-8\pi \beta^2\mu^2\Omega e^{-4\pi(a^2+\beta^2){\cal U}}\int d^\nu
\vec{r}_1 d^\nu \vec{r}_2 (f^{-1}_{\vec{r}_1 \vec{r}_2})^2
[\cosh(8\pi a \beta f^{-1}_{\vec{r}_0 \vec{r}_2})-1]
 \nonumber  \\ &&
+16\pi a \beta \mu^2\Omega e^{-4\pi(a^2+\beta^2){\cal U}}\int d^\nu
\vec{r}_1 d^\nu \vec{r}_2 f^{-1}_{\vec{r}_0
\vec{r}_1}f^{-1}_{\vec{r}_1 \vec{r}_2}\sinh(8\pi a \beta
f^{-1}_{\vec{r}_0 \vec{r}_2})
 \nonumber  \\ &&
-2\Omega^2e^{-4\pi(a^2+2\beta^2){\cal U}}\int d^\nu \vec{r}_1 d^\nu
\vec{r}_2 \cosh(8\pi \beta^2 f^{-1}_{\vec{r}_1 \vec{r}_2})
 \nonumber \\ &&
-4\Omega^2 e^{-4\pi(a^2+2\beta^2){\cal U}}\int d^\nu \vec{r}_1 d^\nu
\vec{r}_2 [\cosh(8\pi a \beta f^{-1}_{\vec{r}_0 \vec{r}_1})-1]
 \nonumber \\ &&
+\Omega^2 e^{-4\pi(a^2+2\beta^2){\cal U}}\int d^\nu \vec{r}_1 d^\nu
\vec{r}_2\Bigl\{e^{-8\pi\beta^2 f^{-1}_{\vec{r}_1
\vec{r}_2}}\cosh[8\pi a \beta (f^{-1}_{\vec{r}_0
\vec{r}_1}+f^{-1}_{\vec{r}_0 \vec{r}_2})]
 \nonumber \\ && \hspace*{5.5cm}
+e^{8\pi\beta^2 f^{-1}_{\vec{r}_1 \vec{r}_2}} e^{-8\pi a \beta
(f^{-1}_{\vec{r}_0 \vec{r}_1}-f^{-1}_{\vec{r}_0 \vec{r}_2})}\Bigl\}
 \;.
\end{eqnarray}
Noting that the coordinates $\vec{r}_1$ and $\vec{r}_2$ have
equivalent positions, one can easily find that Eq.(44) is
identical to Eq.(39). Thus, instead of directly calculating path
integrals, we have reobtained Eqs.(32), (35) and (39) only by
finishing ordinary integrations in the Gaussian smearing formulae.

\subsection{Borrowing Feynman Diagrammatic Technique}
\label{Feynm}

In this subsection, we show that the results Eqs.(32), (35) and
(39) can be reobtained by borrowing Feynman diagrammatic technique
with the propagator Eq.(7). It is well known that one-point
functions in the perturbation theory are the sum of all connected
Feynman diagrams \cite{34}. This point is also true for Eqs.(15)
and (31) because the VP expansion is formally similar to the
perturbative expansion, and accordingly $G_a^{(n)}$ in Eq.(31) is
only the sum of all $n$th-order connected Feynman diagrams
constructed with an external vertex arising from $\exp\{i
\sqrt{8\pi}a \phi(0)\}$ and $n$ internal vertices arising from
${\cal H}_I^{sG}(\phi_{\vec{r}},\mu)$. The Feynman diagrammatic
technique for the sG perturbation field theory has developed early
in 1970s. For example, to perturbatively check the equivalence
between the Coulomb gas and the sG field theory, Feynman
diagrammatic technique was established in 1978 \cite{35}, and
Feynman diagrams in the sG perturbation field theory up to the
third order were also analyzed to consider the renormalization
problem in the sG perturbation theory \cite{36}. In our VP
expansion, the Feynman diagrammatic technique is similar to the
perturbation theory, but, differently, the propagator is Eq.(7),
and vertices are classified as external vertices which come from
the exponential fields and internal vertices which come from
${\cal H}^{sG}_I(\phi_{\vec{r}},\mu)$ in Eq.(30).

Expanding the exponential field $e^{i\sqrt{8\pi}a\phi(0)}$ in
powers of $\phi(0)$, one will find that there exist infinitely
many external vertices (we will call them $e_n$-vertices), having
any number of legs, and the coefficient $e_n\equiv e^{4\pi a^2
I_{(1)}({\cal M}^2)}{\frac {(i\sqrt{8\pi}a)^n}{n!}}$ adheres to
the external $n$-leg vertex, the $e_n$-vertex. Furthermore,
Expanding the cosine field $\cos(\sqrt{8\pi}\beta\phi_{\vec{r}})$
of Eq.(30) in powers of $\phi(\vec{r})$, one can have infinitely
many internal vertices (called $C_{2n}$-vertices here), having any
even number of legs, and an internal $2n$-leg vertex,
$C_{2n}$-vertex, possesses the coefficient $C_{2n}\equiv -2\Omega
e^{4\pi \beta^2 I_{(1)}({\cal M}^2)}{\frac
{(i\sqrt{8\pi}\beta)^{2n}}{(2n)!}}$. Besides, the first term of
Eq.(30) contributes an additional internal $2$-leg vertex (called
$\mu$-vertex) which the coefficient $\mu_2=-{\frac {1}{2}}\mu^2$
adheres to. With all the vertices and the propagator, one can draw
out all the Feynman diagrams order by order for $G_a$ and obtain a
diagrammatic expansion of $G_a$ which is identical to Eq.(31).

The zeroth order diagrams for $G_a$ are all possible connected
graphs self-contracted by the $e_n$-vertices. The external
vertices with $(2n+1)$ legs cannot be completely self-contracted
and have no contributions to $G_a$, which correspond to the
vanishing contributions from the odd powers of the fields. Thus
$G_a^{(0)}$ is the sum of all possible connected graphs
self-contracted by the $e_{2n}$-vertices. We draw them in Fig.1
(Feynman diagrams are drawn with JaxoDraw package \cite{37}).
\begin{figure}[h]
\includegraphics{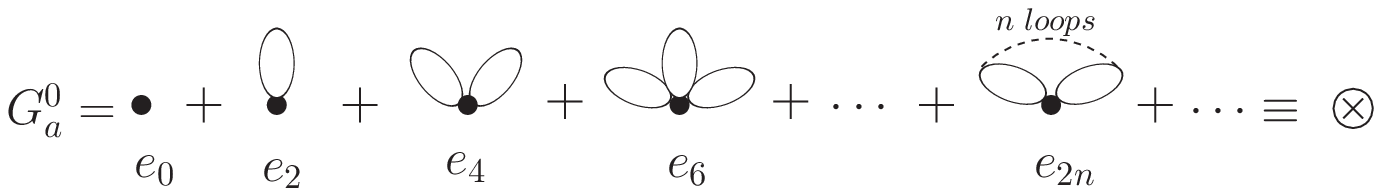}
\caption{\label{fig:1} The Feynman diagrammatic form of $G_a^{(0)}$}
\end{figure}
In Fig.1, an $n$-loop graph has a factor ${\frac
{(2n)!}{n!(2!)^n}}$ which is the number of ways to self-contracted
legs of the $e_{2n}$-vertex. From this graphic form, one can have
\begin{equation}
G_a^{(0)}=\sum_{n=0}^\infty e^{4\pi a^2 I_{(1)}({\cal
 M}^2)}{\frac {(i\sqrt{8\pi}a)^{2n}}{(2n)!}} (f_{00}^{-1})^n \times {\frac
 {(2n)!}{n!(2!)^n}} \;.
\end{equation}
Obviously, Eq.(45) is identical to Eq.(32).

Now we consider the first order diagrams for $G_a$. They are all
possible connected graphs each of which consists of any
$e_n$-vertex and any internal vertex. The diagrams formed by an
external $(2n+1)$-leg vertex and an internal vertex are not
completely contracted because any internal vertex has even number
of legs, and accordingly $G_a^{(1)}$ is the sum of all connected
graphs each of which is constructed by an $e_{2n}$-vertex and an
internal vertex. These graphs have two types: each of one type
consists of an $e_{2n}$-vertex and the $\mu$-vertex
($e_{2n}$-$\mu$ graph) and each of another type consists of an
$e_{2n}$-vertex and a $C_{2n}$-vertex ($e_{2n}$-$C_{2n}$ graph).
We first consider the sum of the $e_{2n}$-$\mu$ graphs. It is
shown in Fig.2.
\begin{figure}[h]
\includegraphics{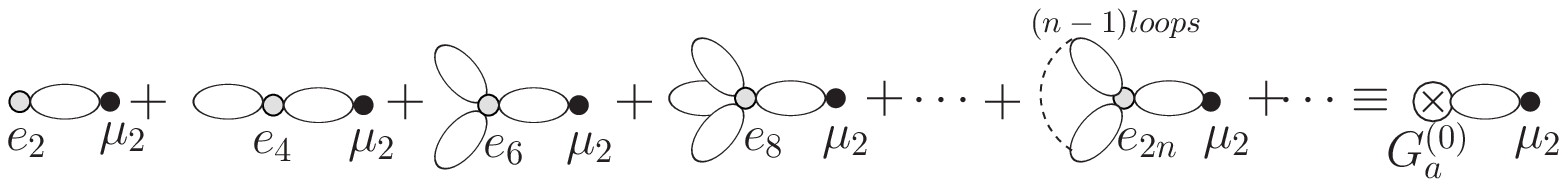}
\caption{\label{Fig.2} The sum of all connected first-order graphs
consisting of $e_{2n}$-vertices and the $\mu$-vertex.}
\end{figure}
In Fig.2, an $e_{2n}$-$\mu$ graph has a factor ${\frac
{(2n)!}{(n-1)!(2!)^{(n-1)}}}$. The sum in Fig.2, $G_{a,I}^{(1)}$,
has the following result
\begin{eqnarray}
G_{a,I}^{(1)}&=&-\sum_{n=1}^\infty (f_{00}^{-1})^{n-1} e^{4\pi a^2
I_{(1)}({\cal M}^2)}{\frac {(i\sqrt{8\pi}a)^{2n}}{(2n)!}}\int d^\nu
\vec{r}_1 (f^{-1}_{0 \vec{r}_1})^2 (-{\frac {1}{2}}\mu^2) \cdot
{\frac {(2n)!}{(n-1)!(2!)^{n-1}}} \nonumber \\ &=&{\frac
{1}{2}}\mu^2(i\sqrt{8\pi}a)^2\int d^\nu \vec{r}_1 (f^{-1}_{0
\vec{r}_1})^2 G_a^{(0)}\;.
\end{eqnarray}
Note that at the first order of $\epsilon$, there is an additional
total factor ${\frac {(-1)^1}{1}}$ the expansion factor, and we
have added it to Eq.(46). We do so in the later Eq.(47), too.

Comparing the result Eq.(46) with Fig.2, one can see that the sum
of all possible connected graphs consisting of the
$e_{2n}$-vertices and the $\mu$-vertex amounts to the sum of all
graphs consisting of an external $2$-leg vertex with a coefficient
${\frac {(i\sqrt{8\pi}a)^2}{2!}}G_a^{(0)}$ (the vertex represented
by the cross-centered circle with two legs in the right hand side
of Fig.2) and the $\mu$-vertex (in fact, only two of such graphs).
Noting that the $\mu$-vertex has two legs, one would admit that
the sum of all possible connected graphs consisting of external
vertices and an $n$-leg internal vertex amounts to the sum of all
graphs consisting of an external $n$-leg vertex with a coefficient
${\frac {(i\sqrt{8\pi}a)^n}{n!}}G_a^{(0)}$ (we will call it an
external $n$-leg exponential vertex) and the $n$-leg internal
vertex (one can prove it directly in the way of getting to
Eq.(46)). Similarly to those in Fig.1 and Fig.2, we will
graphically represent an external $n$-leg exponential vertex by a
cross-centered circle with $n$ legs. Furthermore, one can also
check that the sum of all connected graphs which consist of
$C_{2n}$-vertices and some external or internal $2n$-leg vertex
amounts to summing up all connected graphs which are made by an
internal $2n$-leg vertex with a coefficient $ -2\Omega {\frac
{(i\sqrt{8\pi}\beta)^{2n}}{(2n)!}}e^{-4\pi \beta^2 {\cal U}}$ (we
will call it an internal $2n$-leg cosine vertex) and the external
or internal $2n$-leg vertex. We will graphically represent the
internal $2n$-leg cosine vertex by a dot-centered circle with $2n$
legs.

From the analysis and results in last paragraph, the sum of all
$e_{2n}$-$C_{2n}$ graphs can be depicted in Fig.3.
\begin{figure}[h]
\includegraphics{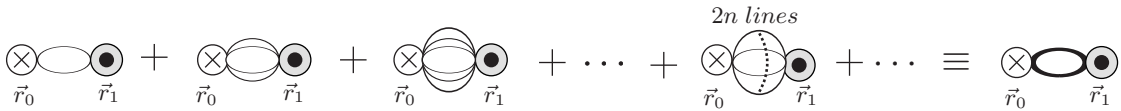}
\caption{\label{Fig.3} The sum of all connected first-order graphs
consisting of $e_{2n}$-vertices and the $C_{2n}$-vertices.}
\end{figure}
In Fig.3, $\vec{r}_0=0$ and a $2n$-line graph has a factor
$(2n)!$. Note that in the right hand side of Fig.3, we have used
the graph with two bold lines to represent the sum of all possible
graphs with even lines between two vertices. Thus, the sum in
Fig.3, $G_{a,II}^{(1)}$, can be calculated as
\begin{eqnarray}
G_{a,II}^{(1)}&=&2\Omega e^{-4\pi \beta^2 {\cal U}}G_a^{(0)}\int
d^\nu \vec{r}_1 \sum^\infty_{n=1} {\frac
{(i\sqrt{8\pi}a)^{2n}}{(2n)!}}(f^{-1}_{0 \vec{r}_1})^{2n}{\frac
{(i\sqrt{8\pi}\beta)^{2n}}{(2n)!}} \cdot (2n)!\nonumber \\
&=&2\Omega e^{-4\pi (a^2+\beta^2) {\cal U}}\int d^\nu
\vec{r}_1[\cosh(8\pi a\beta f^{-1}_{0 \vec{r}_1})-1]\;.
\end{eqnarray}
Comparing last equation and the right hand side of Fig.3, one see
that the two bold lines corresponds to the integrand $[\cosh(8\pi
a\beta f^{-1}_{0 \vec{r}_1})-1]$ and the factor before the
integral in Eq.(47) comes from the $n$-independent parts of the
coefficients for the external $2n$-leg exponential and internal
$2n$-leg cosine vertices. It is evident that summing up
$G_{a,I}^{(1)}$ in Eq.(46) and $G_{a,II}^{(1)}$ in Eq.(47) gives
rise to Eq.(35).

At the second order, diagrams will be complicated. There exist
nine types of graphs: two types for one $e_{2n}$ vertex connecting
with two $\mu$-vertices (figures (a) and (b) in Fig.4), four types
for one $e_{2n}$ vertex connecting with two $C_{2n}$-vertices
(figures (f), (g), (h) and (i) in Fig.4), and three types for
connections among one $e_{2n}$ vertex, one $\mu$-vertex and one
$C_{2n}$-vertex (figures (c), (d) and (e) in Fig.4). The sum of
every type of graphs is depicted in Fig.4.
\begin{figure}[h]
\includegraphics{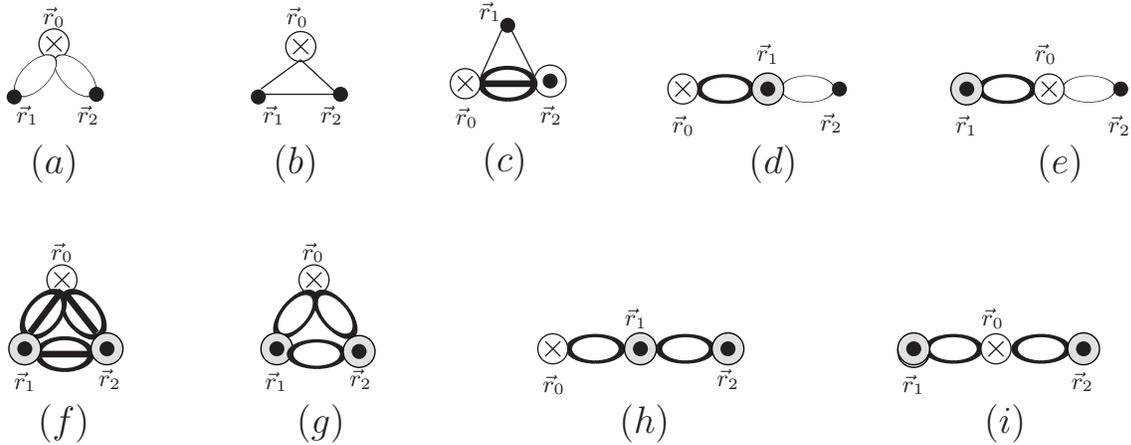}
\caption{\label{Fig.4} The sum of all connected second-order graphs
consisting of $e_{2n}$-, $C_{2n}$-vertices and $\mu$-vertex.}
\end{figure}
In Fig.4, the graph with three bold lines between two vertices
represents the sum of all similar graphs with odd lines between
the same two vertices, and a set of three bold lines corresponds
to a negative hyperbolic sine function (one can check it by doing
as was done in Eq.(47)). The results of graphs in Fig.4 can be
calculated as follow, respectively. The graph $(a)$ in Fig.4 is
simply
\begin{eqnarray}
G_{a,a}^{(2)}&=&{\frac {1}{2}}{\frac
{(i\sqrt{8\pi}a)^4}{4!}}G_a^{(0)}\int d^\nu \vec{r}_1 \int d^\nu
\vec{r}_2 (f^{-1}_{0 \vec{r}_1})^{2}(f^{-1}_{0
\vec{r}_2})^{2}(-{\frac {1}{2}}\mu^2)(-{\frac {1}{2}}\mu^2)\cdot
C_4^2\cdot 2\cdot 2\nonumber \\ &=&8\pi^2 a^4\mu^4 e^{-4\pi a^2
{\cal U}}\int d^\nu \vec{r}_1 \int d^\nu \vec{r}_2 (f^{-1}_{0
\vec{r}_1})^{2}(f^{-1}_{0 \vec{r}_2})^{2}
\end{eqnarray}
In Eq.(48), the symbol $C_4^2$ is a number of combinations, and we
have added the factor ${\frac {1}{2}}$ which is the total factor
at the second order. We also do so for the other second-order
graphs. The graph $(b)$, $(c)$, $(d)$, $(e)$, $(f)$, $(g)$, $(h)$
and $(i)$ in Fig.4 can be written as
\begin{eqnarray}
G_{a,b}^{(2)}&=&{\frac {1}{2}}{\frac
{(i\sqrt{8\pi}a)^2}{2!}}G_a^{(0)}\int d^\nu \vec{r}_1 \int d^\nu
\vec{r}_2 f^{-1}_{0 \vec{r}_1}f^{-1}_{0 \vec{r}_2}f^{-1}_{\vec{r}_1
\vec{r}_2}(-{\frac {1}{2}}\mu^2)(-{\frac {1}{2}}\mu^2)\cdot
C_2^1\cdot 2\cdot 2
 \nonumber \\
&=&-4\pi a^2\mu^4 e^{-4\pi a^2 {\cal U}}\int d^\nu \vec{r}_1 \int
d^\nu \vec{r}_2 f^{-1}_{0 \vec{r}_1}f^{-1}_{0
\vec{r}_2}f^{-1}_{\vec{r}_1 \vec{r}_2} \;,
\end{eqnarray}
\begin{eqnarray}
G_{a,c}^{(2)}&=&{\frac {1}{2}}{\frac
{(i\sqrt{8\pi}a)^1}{1!}}G_a^{(0)}\int d^\nu \vec{r}_1 \int d^\nu
\vec{r}_2 f^{-1}_{0 \vec{r}_1}f^{-1}_{\vec{r}_1
\vec{r}_2}\{-\sinh[8\pi a \beta f^{-1}_{0 \vec{r}_2}]\}
 \nonumber \\&& \hspace*{2cm}
\times (-{\frac {1}{2}}\mu^2) (-2\Omega) {\frac
{(i\sqrt{8\pi}\beta)^{1}}{1!}} e^{-4\pi \beta^2 {\cal U}}
\cdot 2 \cdot 2
 \nonumber \\
&=&16\pi a \beta\mu^2 \Omega e^{-4\pi (a^2+\beta^2) {\cal U}}\int
d^\nu \vec{r}_1 \int d^\nu \vec{r}_2 f^{-1}_{0
\vec{r}_1}f^{-1}_{\vec{r}_1 \vec{r}_2}\sinh[8\pi a \beta f^{-1}_{0
\vec{r}_2}] \;,\hspace*{2cm}
\end{eqnarray}
\begin{eqnarray}
G_{a,d}^{(2)}&=&{\frac {1}{2}}G_a^{(0)}\int d^\nu \vec{r}_1 \int
d^\nu \vec{r}_2 [\cosh(8\pi a\beta f^{-1}_{0 \vec{r}_1})-1]
 \nonumber \\&& \hspace*{2cm}\times
(-2\Omega) {\frac
{(i\sqrt{8\pi}\beta)^{2}}{2!}} e^{-4\pi \beta^2 {\cal
U}}(f^{-1}_{\vec{r}_1 \vec{r}_2})^2(-{\frac
{1}{2}}\mu^2)\cdot 2\cdot 2 \nonumber \\
&=&-8\pi \beta^2\mu^2 \Omega e^{-4\pi (a^2+\beta^2) {\cal U}}\int
d^\nu \vec{r}_1 \int d^\nu \vec{r}_2(f^{-1}_{\vec{r}_1
\vec{r}_2})^2[\cosh(8\pi a\beta f^{-1}_{0 \vec{r}_1})-1]
\;,\hspace*{2cm}
\end{eqnarray}
\begin{eqnarray}
G_{a,e}^{(2)}&=&{\frac {1}{2}}\int d^\nu \vec{r}_1 \int d^\nu
\vec{r}_2 (-2\Omega)e^{-4\pi \beta^2 {\cal U}}[\cosh(8\pi a\beta
f^{-1}_{0 \vec{r}_1})-1]
 \nonumber \\&& \hspace*{2cm}\times
G_a^{(0)} {\frac {(i\sqrt{8\pi} a)^{2}}{2!}} (f^{-1}_{0
\vec{r}_2})^2(-{\frac
{1}{2}}\mu^2)\cdot 2\cdot 2 \nonumber \\
&=&-8\pi a^2\mu^2 \Omega e^{-4\pi (a^2+\beta^2) {\cal U}}\int d^\nu
\vec{r}_1 \int d^\nu \vec{r}_2 (f^{-1}_{0 \vec{r}_2})^2[\cosh(8\pi
a\beta f^{-1}_{0 \vec{r}_1})-1] \;,\hspace*{1.5cm}
\end{eqnarray}
\begin{eqnarray}
G_{a,f}^{(2)}&=&{\frac {1}{2}}G_a^{(0)}\int d^\nu \vec{r}_1 \int
d^\nu \vec{r}_2 (-2\Omega) e^{-4\pi \beta^2 {\cal U}}[-\sinh(8\pi
a\beta f^{-1}_{0 \vec{r}_1})] \hspace*{4cm}
 \nonumber \\&& \hspace*{2cm}\times
(-2\Omega) e^{-4\pi \beta^2 {\cal U}}[-\sinh(8\pi a\beta f^{-1}_{0
\vec{r}_2})][-\sinh(8\pi \beta^2 f^{-1}_{\vec{r}_1 \vec{r}_2})]
 \nonumber \\
&=&-2\Omega^2 e^{-4\pi (a^2+2\beta^2) {\cal U}}\int d^\nu \vec{r}_1
\int d^\nu \vec{r}_2 \sinh(8\pi a\beta f^{-1}_{0 \vec{r}_1})
 \nonumber \\&& \hspace*{3cm}\times
\sinh(8\pi a \beta f^{-1}_{0 \vec{r}_2})\sinh(8\pi \beta^2
f^{-1}_{\vec{r}_1 \vec{r}_2}) \;,\hspace*{2cm}
\end{eqnarray}
\begin{eqnarray}
G_{a,g}^{(2)}&=&{\frac {1}{2}}G_a^{(0)}\int d^\nu \vec{r}_1 \int
d^\nu \vec{r}_2 (-2\Omega) e^{-4\pi \beta^2 {\cal U}}[\cosh(8\pi
a\beta f^{-1}_{0 \vec{r}_1})-1] \hspace*{4cm}
 \nonumber \\&& \hspace*{2cm}\times
(-2\Omega) e^{-4\pi \beta^2 {\cal U}}[\cosh(8\pi a\beta f^{-1}_{0
\vec{r}_2})-1][\cosh(8\pi \beta^2 f^{-1}_{\vec{r}_1 \vec{r}_2})-1]
 \nonumber \\
&=&2\Omega^2 e^{-4\pi (a^2+2\beta^2) {\cal U}}\int d^\nu \vec{r}_1
\int d^\nu \vec{r}_2 [\cosh(8\pi a\beta f^{-1}_{0 \vec{r}_1})-1]
\nonumber \\&& \hspace*{2cm}\times [\cosh(8\pi a\beta f^{-1}_{0
\vec{r}_2})-1][\cosh(8\pi \beta^2 f^{-1}_{\vec{r}_1
\vec{r}_2})-1]\;,\hspace*{2cm}
\end{eqnarray}
\begin{eqnarray}
G_{a,h}^{(2)}&=&{\frac {1}{2}}G_a^{(0)}\int d^\nu \vec{r}_1 \int
d^\nu \vec{r}_2 (-2\Omega) e^{-4\pi \beta^2 {\cal U}}[\cosh(8\pi
a\beta f^{-1}_{0 \vec{r}_1})-1]
 \nonumber \\&& \hspace*{6cm}\times
(-2\Omega) e^{-4\pi \beta^2 {\cal U}}[\cosh(8\pi \beta^2
f^{-1}_{\vec{r}_1 \vec{r}_2})-1]
 \nonumber \\
&=&2\Omega^2 e^{-4\pi (a^2+2\beta^2) {\cal U}}\int d^\nu \vec{r}_1
\int d^\nu \vec{r}_2 [\cosh(8\pi a\beta f^{-1}_{0 \vec{r}_1})-1]
 \nonumber \\&& \hspace*{6cm}\times
[\cosh(8\pi \beta^2 f^{-1}_{\vec{r}_1
\vec{r}_2})-1]\;,\hspace*{2.2cm}
\end{eqnarray}
\begin{eqnarray}
G_{a,i}^{(2)}&=&{\frac {1}{2}}\int d^\nu \vec{r}_1 \int d^\nu
\vec{r}_2 (-2\Omega) e^{-4\pi \beta^2 {\cal U}}[\cosh(8\pi a\beta
f^{-1}_{0 \vec{r}_1})-1]
 \nonumber \\&& \hspace*{4cm}\times
G_a^{(0)}(-2\Omega) e^{-4\pi \beta^2 {\cal U}} [\cosh(8\pi a\beta
f^{-1}_{0 \vec{r}_2})-1]
 \nonumber \\
&=& 2\Omega^2 e^{-4\pi (a^2+2\beta^2) {\cal U}}\{\int d^\nu
\vec{r}_1 [\cosh(8\pi a\beta f^{-1}_{0 \vec{r}_1})-1]\}^2
\;.\hspace*{4cm}
\end{eqnarray}
For easily reading, every first expression in Eqs.(48)---(56) was
written down in the way of one factor in the expression
corresponding to one part in the relevant graph (some graphs have
symmetrical factors). One can check that the sum of
Eqs.(48)---(56) coincides with Eq.(39). Thus, one has seen that
the VP expansion of one-point functions can be performed by
borrowing Feynman diagrammatic technique in conventional
perturbative theory.

\section{sG and shG Field Theories with $\nu=2$: One-order VP Results}
\label{4}

Subsection A and B will provide comparisons of the one-order VP
results for sG and shG field theories with the conjectured exact
results, respectively, and subsection C will compare the one-order
VP results with the one-order perturbative results sG field
theories .

\subsection{Comparisons with the Conjectured Results: sG Field Theory}
\label{sG}

In two-dimensional Euclidean space, Eqs.(32) and (35) give
($\epsilon=1$)
\begin{eqnarray*}
G_a^{I}&=&({\frac {\mu^2}{{\cal M}^2}})^{a^2}-4\pi \mu^2 a^2
({\frac {\mu^2}{{\cal M}^2}})^{a^2}\int d^2 \vec{r}_1 (f^{-1}_{0
\vec{r}_1})^2 \nonumber  \\ &&\ \  \ \ \ +2 \Omega ({\frac
{\mu^2}{{\cal M}^2}})^{a^2+\beta^2}\sum_{n=1}^\infty {\frac {(8\pi
a \beta)^{2 n}}{(2 n)!}}\int d^2 \vec{r}_1 (f^{-1}_{0
\vec{r}_1})^{2n}
\end{eqnarray*}
\begin{eqnarray}
\hspace*{1.5cm}= ({\frac {\mu^2}{{\cal M}^2}})^{a^2}(1-2 a^2
K_{02})+4\pi {\frac {\Omega}{{\cal M}^2}} ({\frac {\mu^2}{{\cal
M}^2}})^{a^2+\beta^2-1} K_{0c} \;,
\end{eqnarray}
 where, $$K_{02}\equiv \int^\infty_0 dx x K_0^2(x),  \
\  \  \  \  \ \  \ K_{0c}\equiv \int^\infty_0 dx x [\cosh(4 a \beta
K_0 (x))-1].$$

To determine the auxiliary parameter $\mu$, as stated in Sect. II,
we can require that ${\frac {d G_a^{I}}{d(\mu^2)}}=0$ according to
the PMS, and have
\begin{equation}
{\frac {\mu^2}{{\cal M}^2}}=\biggl(4 \pi {\frac {\Omega}{{\cal
M}^2}} K_{0c} {\frac {1-a^2-\beta^2}{a^2(1-2 a^2
K_{02})}}\biggl)^{1/(1-\beta^2)} \;.
\end{equation}
Thus, Eq.(57) with Eq.(58) gives $G_a^I$,  the approximate result
of $G_a$ up to the first order in the VP theory. Note that the
normal-ordering parameter ${\cal M}$ can be taken as any positive
value with mass dimension, and different choice of its value leads
to just a finite multiplicative redefinition of $\Omega$ \cite{25}
(1975). Once ${\cal M}$ is chosen, the theory is defined and the
parameter $\Omega$ is given a precise meaning, just as the choice
of the normalization of the field $\cos(\beta\phi)$ in
Ref.~\cite{4} does. We will take ${\cal M}$ as unit. This choice
renders every quantities in Eqs.(57) and (58) dimensionless, and
simultaneously amounts to taking the same normalization conditions
as Eqs.(6) and (16) in Ref.~\cite{4}. The latter point can be
checked by directly calculating the relevant two-point correlation
function with the two points approaching each other within our
formalism and then comparing the result with that in
Ref.~\cite{4}. Thus, our results can be compared with the exact
formula. Because being exponential-like interaction, the
$two$-dimensional sG field theory can be rendered finite only for
the case $\beta^2<{\frac {1}{2}}$ by the Coleman's normal-ordering
prescription (it amounts to the renormalization of the mass
parameter. For the range of $\beta$ where the sG field theory
diverges, one has to resort to a further renormalization
procedure), and accordingly $G_a$ is finite for the case
$\beta^2<{\frac {1}{2}}$ and $a\beta<{\frac {1}{2}}$. And so
$G_a^{I}$ here is finite only for the range $a\beta<{\frac
{1}{2}}$ (one can see it by noticing the integral $K_{0c}$).
Fortunately, this range of $\beta$ is just the validity scope of
the conjectured exact formula in Ref.~\cite{4}. Numerically,
$G^{I}_a$ can be given with Mathematica programme for the range of
$\alpha \beta< 0.426925$, as stated in Ref.~\cite{27}. This also
occurs at the shG field theory.

Now we consider the comparison for the sG field theory. From
Eq.(12) in Ref.~\cite{4}, the dimensionless $\Omega$ can be
written as
\begin{equation}
\Omega={\frac {\Gamma(\beta^2)}{\pi \Gamma(1-\beta^2)}}\biggl[{\frac
{\sqrt{\pi}\Gamma({\frac {1}{2}}+{\frac {\xi}{2}})}{2 \Gamma({\frac
{\xi}{2}})}}\biggl]^{2-2\beta^2}
\end{equation}
with $\xi={\frac {\beta^2}{1-\beta^2}}$. In last equation, we took
the soliton mass as unit to make $\Omega$ dimensionless. For the
conjectured exact formula Eq.(20) in Ref.~\cite{4}, we also do so
and have
\begin{eqnarray}
G_a^{exact}&=&{\frac {2 \sin({\frac {\pi\xi}{2}})\Gamma({\frac
{1}{2}}+{\frac {\xi}{2}})\Gamma(1-{\frac {\xi}{2}})}{4\sqrt{\pi}}}
\nonumber \\ && \hspace*{1cm}\exp\{\int^\infty_0{\frac
{dt}{t}}[{\frac {\sinh^2(2 a \beta t)}{2 \sinh(\beta^2
t)\sinh(t)\cosh((1-\beta^2)t)}}-2 a^2 e^{-2t}]\} \;.
\end{eqnarray}
To compare $G_a^{I}$ with the conjectured exact result, we depict
$G_a^{I}$ according to Eq.(57) with Eqs.(58) and (59) as the left
figure of Fig.5, and the exact result $G^{exact}_a$ according to
Eq.(60) as the right figure of Fig.5.
\begin{figure}[h]
\includegraphics{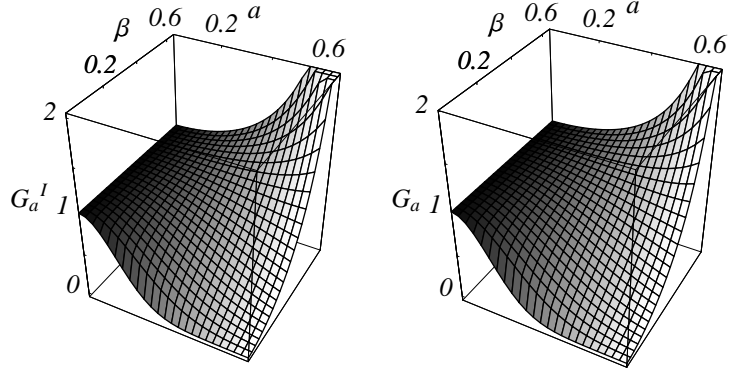}
\caption{\label{Fig.5} Comparison of $G_a^{I}$ (the left) with the
conjectured exact result $G_a^{exact}$ (the right).}
\end{figure}
In Fig.5, the parameter range is $0\le a<0.65$ and  $0\le
\beta<0.65$. In principle, the two figures in Fig.5 can be
extended to all range $\alpha \beta< 0.5$. From Fig.5, one can see
that for smaller $\beta$ and larger $a$, $G_a$ tends to zero, and
for larger $\beta$ and $a$, $G_a$ is greater than the value $1$.
When a large $\beta$ is given, $G_a$ increases with the increase
of $a$, and when a small $\beta$ is given, $G_a$ decreases with
the increase of $a$. On the other hand, the two figures in Fig.5
resemble each other very well and suggest that for the range of
the parameter $\{a,\beta\}$ drawn in Fig.5 $G_a^{I}$ has a good
agreement with the conjectured exact result. Our numerical
analysis indicates that for the case of $a<0.2$ or so, the
one-order results almost completely agree with the conjectured
exact results, and for larger $a$, our results differ from the
exact results with about ten percents or so (at most with 20 more
percents when $a$ approaches values with $a \beta\le 0.426925$
satisfied, see Table I in Ref.~\cite{27}).
\begin{figure}[h]
\includegraphics{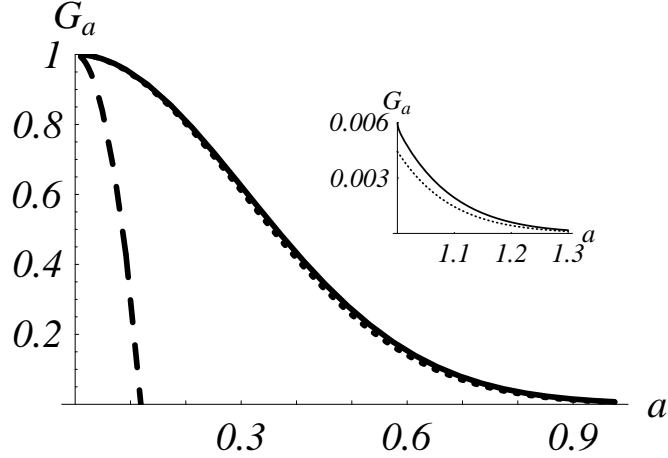}
\caption{\label{fig.6} Comparison between the one-order ($G_a^I$,
solid curve) and conjectured exact ($G_a^{exact}$, dotted curve)
sG VEVs of exponential fields at $\beta=0.15$. The dashed curve is
the one-order perturbative result ($G_a^{Ipert}$, see subsection
C) at the same value of $\beta$. At this value of
$\beta$,$G_a^{Ipert}$ and $G_a^{exact}$ are widely discrepant.}
\end{figure}
\begin{figure}[h]
\includegraphics{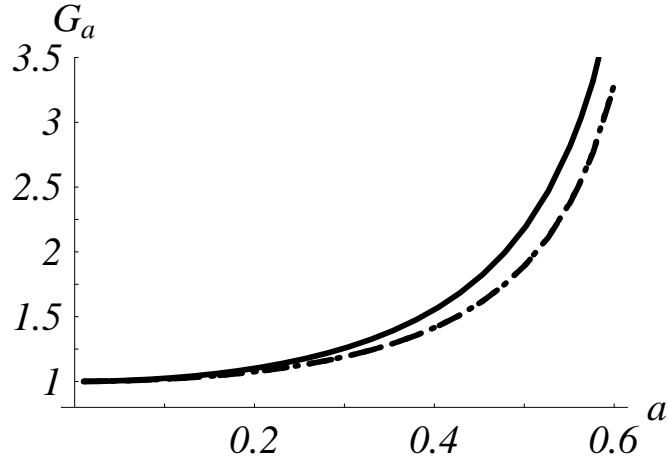}
\caption{\label{fig.7} Similar to Fig.6 but at $\beta=0.7$. At
this case, the one-order perturbative result (the dashed curve) is
almost coincident with the conjectured exact result.}
\end{figure}
For a concrete illustration, we draw $G_a^I$ (the solid curves)
and the exact result $G_a^{exact}$ (the dotted curves) at
$\beta=0.15$ and $0.7$ in Fig.6 and Fig.7, respectively (the
dashed curves is the perturbative results, see next subsection C).
These two figures indicate that $G_a^{I}$ is always greater than
$G_a^{exact}$.

\subsection{Comparisons with the Conjectured Results: shG Field Theory}
\label{shG}

In the Euclidean space, the shG Lagrangian density is
\begin{equation}
{\cal L}_{shG}= {\frac {1}{2}}\nabla_{\vec{r}}\phi_{\vec{r}}
\nabla_{\vec{r}} \phi_{\vec{r}} + 2
\Delta\cosh(\sqrt{8\pi}\gamma\phi_{\vec{r}})\;.
\end{equation}
For the convenience of comparison with the conjectured exact VEV
of the exponential field in the shG field theory in Ref.~\cite{5},
here we are interested in the VEV $G_b\equiv <e^{\sqrt{8\pi} b
\phi(0)}>$. Taking $a=-ib$, $\beta=i\gamma$ and $\Omega=-\Delta$
in Eq.(57), we get the one-order VP approximate result of $G_b$,
$G_b^{I}$, as follows
\begin{equation}
G_b^{I}= (\mu^2)^{-b^2}(1+2 b^2 K_{02})-4\pi \Delta
(\mu^2)^{-b^2-\gamma^2-1} K_{0ch}
\end{equation}
with the parameter $\mu^2$ satisfied
\begin{equation}
\mu^2=\biggl(4 \pi \Delta K_{0ch} {\frac {1+b^2+\gamma^2}{b^2(1+2
b^2 K_{02})}}\biggl)^{1/(1+\gamma^2)} \;,
\end{equation}
where, $K_{0ch}\equiv \int^\infty_0 dx x [\cosh(4 b \gamma K_0
(x))-1]$ is similar to $K_{0c}$. From Eqs.(9) and (8) in
Ref.~\cite{5}, the dimensionless $\Delta$ has the following form
\begin{equation}
\Delta=-{\frac {\Gamma(-\gamma^2)}{\pi
\Gamma(1+\gamma^2)}}\biggl[{\frac {\Gamma({\frac {1}{(2+2
\gamma^2)}}) \Gamma(1+{\frac {\gamma^2}{2+2
\gamma^2}})}{4\sqrt{\pi}}}\biggl]^{2+2\gamma^2} \;,
\end{equation}
and the dimensionless conjectured exact expression of $G_b$,
$G_b^{exact}$, is
\begin{eqnarray}
G_b^{exact}&=&\biggl[{\frac {\Gamma({\frac {1}{(2+2 \gamma^2)}})
\Gamma(1+{\frac {\gamma^2}{2+2 \gamma^2}})}{4\sqrt{\pi}}}\biggl]^{-2
b^2} \nonumber
\\ && \hspace*{1cm}\exp\Bigl\{\int^\infty_0{\frac {dt}{t}}\bigl[{\frac
{\sinh^2(2 b \gamma t)}{2 \sinh(\gamma^2
t)\sinh(t)\cosh((1+\gamma^2)t)}}+2 b^2 e^{-2t}\bigl]\Bigl\} \;.
\end{eqnarray}
\begin{figure}[h]
\includegraphics{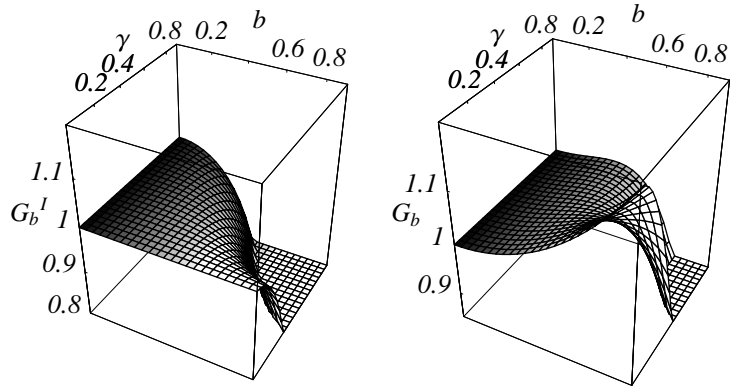}
\caption{\label{Fig.8} Comparison of $G_b^{I}$ with the
conjectured exact $G_b$.}
\end{figure}
In Eqs.(64) and (65), we adopted symbols here and took the
particle mass as unit. Now we can compare $G_b^{I}$ with
$G_b^{exact}$. We depict $G_b^{I}$ (the left figure) and
$G_b^{exact}$ (the right figure) in Fig.8 for the range of $b<0.9$
and $\gamma<0.9$ (it can be extended to the all tractable range of
$b\gamma<0.426925$ ). In the left figure of Fig.8, $G_b^{I}$ is
set to zero for the range of $b\gamma>0.426925$ where Mathematica
couldn't produce finite numerical results. The two surfaces in
Fig.8 are basically resemble each other, except for the cases of
larger $b$ and simultaneously smaller $\gamma$. Numerical analysis
indicates that for a given value of $\beta$, $G_b^{I}$ always
decreases with the increase of $b$, while $G_b^{exact}$ is first
increase and then goes down with the increase of $b$. But, for not
large $b$, the relative differences of $G_b^{I}$ from the
conjectured exact result $G_b^{exact}$ are small, the situations
on the differences between $G_b^{I}$ and $G_b^{exact}$ are
analogous to those for the sG field theory.
\begin{figure}[h]
\includegraphics{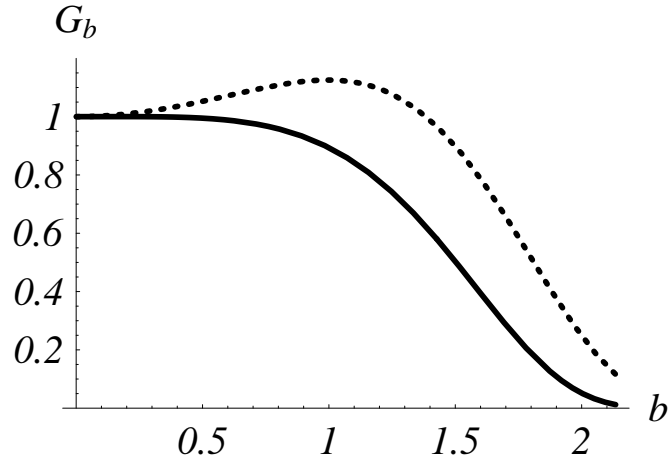}
\caption{\label{Fig.9} Comparison of $G_b^{I}$ (solid) with the
conjectured exact $G_b$ (dotted) at $\gamma=0.2$.}
\end{figure}
\begin{figure}[h]
\includegraphics{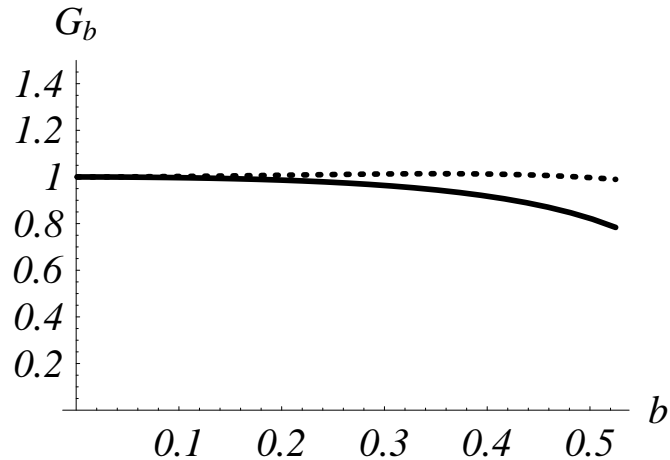}
\caption{\label{Fig.10} Comparison of $G_b^{I}$ (solid) with the
conjectured exact $G_b$ (dotted) at $\gamma=0.8$.}
\end{figure}
We show these points in Fig.9 and Fig.10 for the values of
$\gamma=0.2$ and $\gamma=0.8$, respectively. In Figs.9 and 10, the
dotted curves are for $G_b^{exact}$ and the solid ones for
$G_b^{I}$. Specially, when $b$ is smaller than $0.2$ or so,
$G_b^{I}$ almost completely coincides with the conjectured result
$G_b^{exact}$. These two figures show that for the shG field
theory, $G_b^{I}$ is always smaller than $G_b^{exact}$, which is
opposite to that for the sG field theory.

By the way, if we choose $\Delta$ as unit instead of setting the
particle mass as unit as did in the above, then the numerical
analysis indicates that the dependence of $G_b^{I}$ upon $b$ at a
given $\beta$ is qualitatively similar to that of $G_b^{exact}$.

\subsection{Comparisons with the One-order Perturbative Results: sG Field Theory}
\label{pert}

Using the massive Thirring model, Ref.~\cite{12}(Poghossian)
calculated $G_a$ perturbatively up to the first order of the
coupling $g$, and obtained
\begin{eqnarray}
G_a^{Ipert}&=&({\frac {1}{2}})^{2 a^2}\exp\{\int^\infty_0 {\frac
{dt}{t}}[{\frac {\sinh^2(\sqrt{2} a t)}{\sinh^2(t)}}-2 a^2 e^{-2
t}]\}
 \nonumber  \\ && \hspace*{1cm} \times
[1+{\frac {g}{8\pi}}[-2\Psi({\frac {1}{2}})+8 a^2(1-4 \log2) + 8
a^2(\Psi(\sqrt{2} a)+\Psi(-\sqrt{2} a))
 \nonumber  \\ && \hspace*{2cm}
+(1-8 a^2)(\Psi({\frac {1+2\sqrt{2} a}{2}})+\Psi({\frac
{1-2\sqrt{2} a}{2}}))]] +o(g^2) \;,
\end{eqnarray}
where, the coupling $g$ is related to the sG coupling $\beta$ as
follows
\begin{equation}
{\frac {g}{\pi}}={\frac {1}{2\beta^2}}-1  \,.
\end{equation}
Eq.(67) indicates that $g\to 0$ corresponds to $\beta^2\to {\frac
{1}{2}}$.
\begin{figure}[h]
\includegraphics{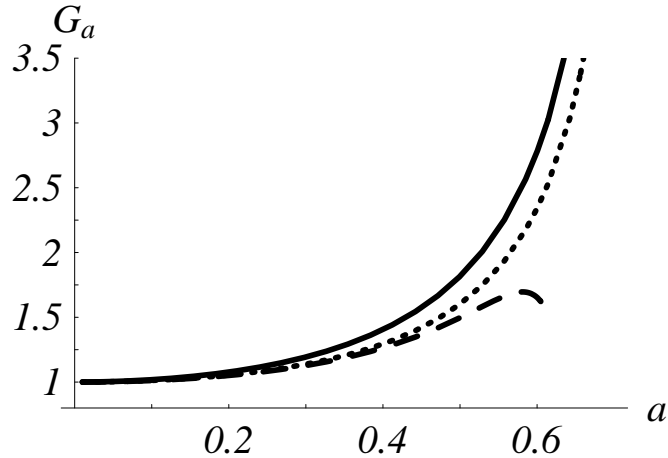}
\caption{\label{Fig.11} Similar to Fig.6 and Fig.7, but at
$\beta=0.65$.}
\end{figure}
In Fig.11, we compare the one-order perturbative (the dashed
curves) and VP (solid) results with the conjectured exact results
(dotted) at the case of $\beta=0.65$. In subsection A of this
section, the same comparisons were made at $\beta=0.15$ and $0.7$
in Fig.6 and Fig.7, respectively. In Figs.7, $\beta^2=0.49$,
$i.e.$, $\beta$ approaches ${\frac {1}{2}}$ or $g\to 0$, the
one-order perturbative results are almost completely identical to
the conjectured exact results for all the plotted values of $a$,
while the one-order VP results have evident differences (the
relative errors are 20\% or so at most) from the conjectured exact
results for larger values of $a$. In Fig.11, $\beta^2\approx
0.42$, the one-order perturbative results (dashed) have quite
large deviations from the conjectured exact results (dotted) for
larger values of $a$, and differences of the one-order VP results
(solid) from the conjectured exact results for larger values of
$a$ get smaller than the case of $\beta^2=0.49$. In Fig.6,
$\beta^2=0.02$, the one-order perturbative results and the
conjectured exact results are widely discrepant for all the
plotted values of $a$, while differences of the one-order VP
results from the conjectured exact results for larger values of
$a$ get much smaller than the cases of $\beta^2=0.49$ and $0.42$.
Besides, in Figs. 6, 7, and 11, the one-order VP results almost
coincide with the conjectured exact results for not large $a$
($<0.2$ or so). To illustrate the dependency of the deviations of
the one-order perturbative and VP results from the conjectured
exact results upon the coupling $\beta$, $G_a^{Ipert}$, $G_a^{I}$
and $G_a^{exact}$ with $a=0.2, 0.3$ and $0.5$ are depicted as
functions of $\beta$ in Figs.12, 13 and 14, respectively.
 \begin{figure}[h]
\includegraphics{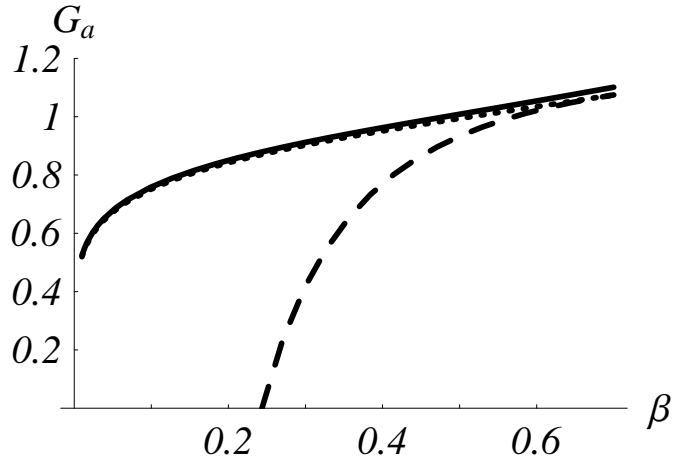}
\caption{\label{Fig.12} The dependances of $G_a^{Ipert}$(dashed),
$G_a^{I}$(solid) and $G_a^{exact}$(dotted) upon $\beta$ for
$a=0.2$.}
\end{figure}
\begin{figure}[h]
\includegraphics{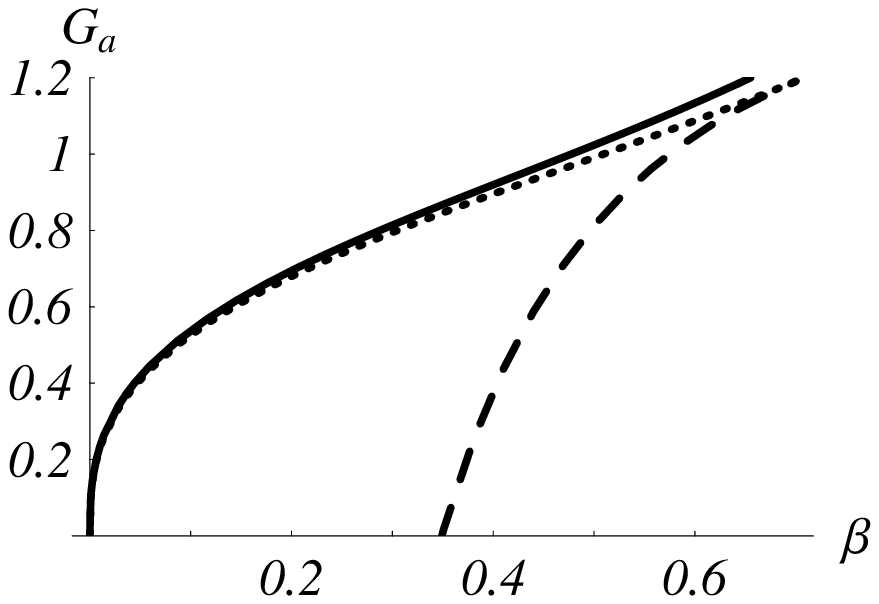}
\caption{\label{Fig.13} The dependances of $G_a^{Ipert}$(dashed),
$G_a^{I}$(solid) and $G_a^{exact}$(dotted) upon $\beta$ for
$a=0.3$.}
\end{figure}
\begin{figure}[h]
\includegraphics{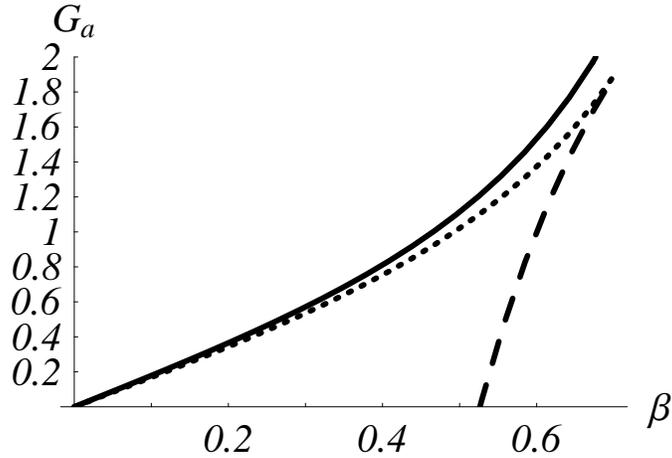}
\caption{\label{Fig.14} The dependances of $G_a^{Ipert}$(dashed),
$G_a^{I}$(solid) and $G_a^{exact}$(dotted) upon $\beta$ for
$a=0.5$.}
\end{figure}
These three figures  evidently show that the one-order
perturbative results (dashed) have good agreements with the
conjectured exact results (dotted) only when $\beta^2$ approaches
the value ${\frac {1}{2}}$ ($i.e.,$ when $g\to 0$), but their
deviations from the conjectured exact results appear, for a given
value of $a$ (not too small), and become larger and larger when
$\beta^2$ decreases from ${\frac {1}{2}}$. For a given $a$, when
values of $\beta^2$ is much smaller than ${\frac {1}{2}}$, the
one-order perturbative and conjectured exact results are widely
discrepant. On the other hand, the one-order VP results (solid)
are almost identical to the conjectured exact results for not
large values of $\beta$, and have not too larger differences from
the conjectured exact results even when $\beta^2$ approaches
${\frac {1}{2}}$.

\section{Conclusion}
\label{5}

In this paper, we developed a VP scheme to calculate the VEVs of
local fields in relativistical QFTs. For a class of scalar field
theories whose potential have Fourier representations, we obtained
the Gaussian smearing formulae for the VEVs of a comparatively
general local field. As an application and illustrations on the
scheme and the Gaussian smearing formulae, we considered the sG
and shG field theories. This application provided an example in
both directly and Feynman-diagrammatically performing the VP
scheme, and showed the usefulness of the Gaussian smearing
formulae. The Gaussian smearing formulae can relieve us of hard
labor in path integrals. The usefulness of the Gaussian smearing
formulae can also be envisioned from the smearing formulae for the
higher order effective classical potential in statistical
mechanics \cite{38}, which succeeded in applying to the singular
Coulomb potential. Additionally, according to the saying in
Ref.~\cite{2} when discussing normal-ordering prescription, the
existence of the Fourier representation of a given $V(\phi)$ for
the validity of the Gaussian smearing formulae is irrelevant in
the derivation of section II, because the final Gaussian smearing
formulae are purely algebraic \cite{2}. Although the one-order VP
results are the lowest approximate in VP theory, the numerical
discussions in last section indicated that the one-order VP VEVs
of the exponential fields for the sG and shG field theories give a
strong support to the conjectured formulae on the exact VEVs in
Refs.~\cite{4,5} and the first paper in Ref.~\cite{6}, at least,
for not large values of the coupling and the exponential-field
parameter. They also suggest the effectiveness of the VP scheme
here in calculating the VEVs of local fields for QFTs. The
comparisons in last section also illustrates the
non-perturbability of the VPT scheme here and its advantages over
the perturbative theory. We believe that the VP scheme in the
present paper can provide an effective, systematical controllable
non-perturbative approximate tool for calculating VEVs of local
fields in relativistic QFTs.

As was pointed out in Ref.~\cite{27}, there exist some interesting
problems based on our work, and here we do not repeat them one by
one. Nevertheless, we intend to stress that a further
investigation on the higher order results would improve the
one-order VP results here and show the convergency of the VP
theory. As a matter of fact, a numerical analysis on the VP
results up to the second order in the sG field theory in
Ref.~\cite{27} has suggested this point. The applications of the
scheme in the present paper to other field theories maybe also
give a good check and substantial support to the other existed
conjectured exact formulae \cite{5,6,7,8,9,10,11,12}.
Simultaneously, the problem of calculating VEVs of local fields
provides a good laboratory for the VP theory because there existed
so many conjectures on VEVs of local fields in various perturbed
conformal field theories. Of course, generalizing the VP scheme
here to other physical problems will be interesting and useful.
For example, the bound state problems in relativistic QFT is a
notoriously difficult non-perturbative problems, and we noticed
that there had existed an approach of attacking it which is a
combination of variational method and perturbative theory
\cite{41}. The approach in Ref.~\cite{41} can be regarded as a
method of calculating VEV, but, as pointed by the authors, it is
valid only for the weak coupling, because the variational
procedure in it is to choose the appropriate two-particle operator
and the expansion scheme is a naive perturbative expansion. We
think that based on the work in this reference, the VP scheme here
can be generalized for calculating bound-state mass, and this
generalization will be meaningful and interesting.

\begin{acknowledgments}
I acknowledge Prof. C. F. Qiao for his drawing my attention on the
Axodraw package. This project was sponsored by SRF for ROCS, SEM
and supported by the National Natural Science Foundation of China
as well as High Performance Computing Center of Shanghai Jiao Tong
University.
\end{acknowledgments}

\end{document}